\documentclass[lettersize,journal]{IEEEtran}
\usepackage{svg}

\usepackage{amsmath,amsfonts,amssymb}
\usepackage{bm}
\usepackage{braket}
\usepackage{gensymb}

\usepackage{mhchem} 
\usepackage{subcaption}
\usepackage{amssymb}
\usepackage{graphicx} 
\usepackage{subcaption} 
\usepackage{amsthm}
\usepackage{amsmath,amssymb}
\usepackage{braket}
\usepackage{gensymb}
\usepackage{multirow}
\usepackage{booktabs}
\usepackage{adjustbox}
\usepackage{array}
\usepackage{bm}
\usepackage{colortbl}
\usepackage{xcolor}
\usepackage{tabularx}
\usepackage{multirow}
\usepackage{multicol}
\usepackage{lipsum} 

\usepackage[noadjust]{cite}
\usepackage[table]{xcolor} 
\usepackage{booktabs}
\usepackage{multirow}
\usepackage{array}
\usepackage{makecell}
\usepackage{adjustbox}
\usepackage{colortbl}
\usepackage{hhline}
\usepackage{threeparttable}

\usepackage{graphicx}
\usepackage{tikz}
\usepackage{subcaption} 
\usepackage{algorithm}
\usepackage[noend]{algpseudocode}

\usepackage{url}
\usepackage{hyperref}
\usepackage{cleveref}
\usepackage{microtype}
\usepackage{underscore}

\begin{document}



\title{Learning Physically Grounded Geometries as Actionable Interpretability for Molecule–Surface Systems via Multimodal Information Bottlenecks}

\title{Learning Physically Meaningful and Actionable Key Geometries via Multimodal Information Bottlenecks for Molecule--Surface System Discovery}

\author{
Zhenzhong Wang,~\IEEEmembership{Member, IEEE},
Nianze Wu,
Chenggong Huang,
Lu Qiu,
Min Jiang$^{\star}$,~\IEEEmembership{Fellow, IEEE}\\

\IEEEcompsocitemizethanks{\IEEEcompsocthanksitem

Zhenzhong Wang, Nianze Wu, Chenggong Huang, Lu Qiu, and Min Jiang are with the School of Informatics, Xiamen University, Xiamen 361005, Fujian, P.R. China. (e-mail: zhenzhongwang@xmu.edu.cn; 19020242202603@stu.xmu.edu.cn; 
huangchenggong@stu.xmu.edu.cn;
luqiu@stu.xmu.edu.cn; minjiang@xmu.edu.cn). \textit{($\star$Corresponding author: Min Jiang.)}

	}	
}



\maketitle

\begin{abstract}
Molecule–surface systems, in which molecules interact with material surfaces, are central to catalysis design and energy conversion. Unlike isolated molecules, the properties of such systems are governed by subtle surface geometries.
Machine learning surrogates for predicting such systems' properties are computationally efficient, but they operate as black-box regressors: they fit target properties without exposing these subtle geometries that govern them. 
On the other hand, existing graph explanation methods are post-hoc: they merely reveal which local geometries a predictor's output relies on. However, in scientific prediction, we should go beyond understanding model outputs and ask a more demanding question: \emph{how can we make the predictor rely on the physically plausible explanations for more reliable predictions?} This raises two challenges: (i) how to make the identified explanations actionable for prediction, and (ii) how to ensure that the identified explanations reflect physically meaningful geometries.
To address these challenges, we propose \emph{Geomo} to learn the underlying structure–property relationships by identifying physically meaningful key motifs for surface-bound molecular system property prediction.
Specifically, {\em Geomo} formulates the motif discovery as a multimodal information bottleneck problem, aiming to identify the motifs that maximally correlate with the target property while closely aligning with textual expert-crafted descriptors.
These motifs are not merely post-hoc explanations. Rather, they serve as physical priors built into the model's reasoning pathway to enhance prediction accuracy and reliability. 
In catalytic configurations, a representative class of molecule–surface systems, \emph{Geomo} achieves competitive performance in energy prediction across diverse out-of-distribution splits and performs well in high-throughput screening for optimal-configuration discovery.
Beyond accuracy, the motifs uncovered by {\em Geomo} quantitatively correlate with charge density maps, confirming that the model learns physically meaningful structure–property relationships. 
\end{abstract}

\begin{IEEEkeywords}
Graph neural networks explainability, multimodal learning, molecular representation learning 
\end{IEEEkeywords}

\section{Introduction}
Molecule–surface systems are interfacial systems in which molecular species are bound to material surfaces. For example, in catalysis, an adsorbate bound to a material surface forms such a system, whose binding energy determines the catalytic activity~\cite{wang2024crystalline,sabatier1920catalyse,medford2015sabatier,chen2024unusual}. Accurately computing the properties of such systems is essential for catalysis design and energy conversion. Unlike isolated molecules, the properties of these systems are governed by subtle surface 3D structures, requiring precise electronic structure calculations to resolve. Traditional workflows rely on computationally expensive first-principles calculations such as Density Functional Theory (DFT). However, their prohibitive computational cost severely limits high-throughput evaluation~\cite{hohenberg1964inhomogeneous,kohn1965self,norskov2011density}. This computational bottleneck has motivated substantial efforts toward efficient data-driven surrogates~\cite{kolluru2024adsorbdiff,yao2024property,11123751,10595522,bozal2025developing,xu2025efficient,luo2026towards,11200037}.

Existing data-driven surrogates for molecule–surface systems fall into three mainstream categories by their used data representations. The first relies on geometric neural networks, which model atomic systems as graphs to encode local geometries and bonding environments~\cite{gilmer2017neural,xie2018crystal,joshi2025all,lan2023adsorbml,10494027}. Classic architectures, including SchNet~\cite{schutt2018schnet}, DimeNet/DimeNet++~\cite{gasteiger2020fast,gasteiger2020directional}, and GemNet~\cite{gasteiger2021gemnet}, adopt distance- and angle-aware message passing to capture fine-grained structural interactions, while EquiformerV2~\cite{liao2023equiformerv2} advances geometric reasoning via equivariant self-attention.
The second category adopts large language-based methods, which encode catalytic adsorption systems as textual sequences~\cite{chen2025multi,ock2023catalyst,ock2024multimodal,wang2026endowingmolecularlanguagegeometry,ock2026adsorb}. As a representative example, CatBERTa~\cite{ock2023catalyst} casts catalytic adsorption energy prediction as a language modeling task and incorporates domain knowledge via pretrained textual representations.
The third emerging direction integrates geometry and text into multimodal frameworks to leverage their complementary merits~\cite{chen2025multi,ock2024multimodal}. Methods such as GAP have validated that joint modeling of structural and semantic information can further boost prediction accuracy.


Despite these advances, these surrogates typically operate as black-box regressors and thus fail to capture subtle structure--property correlations. Understanding such correlations requires identifying the key local geometric arrangement at the surface, because these local environments govern charge redistribution determining adsorption behavior~\cite{wang2024comprehensive,wu2023chemistry,wang2026explainablemolecularpropertyprediction,henderson2021improving,esterhuizen2022interpretable,xin2023interpretable}. Although numerous graph neural network (GNN) explanation methods have been proposed to identify the key structures, such as GNNExplainer~\cite{ying2019gnnexplainer}, MatchExplainer~\cite{pmlrv202wu23jExplaining}, and PGExplainer~\cite{luo2020pgexplainer}, the explanations they produce merely reveal which local geometries a predictor's output relies on. In scientific prediction, however, we should go beyond understanding model outputs and ask a more demanding question: \emph{how can we make the predictor rely on the physically plausible explanations for more reliable predictions?}

To answer this question, two challenges need to be addressed: (i) \emph{how to make the identified explanations actionable for prediction}, and (ii) \emph{how to ensure that the identified explanations reflect physically meaningful local geometries.}
First, actionable explanations should be human-understandable. More importantly, they should be usable by the predictor itself during training, rather than merely produced after training without improving prediction performance. Second, a physically meaningful explanation should not be merely predictive. More importantly, it should reflect expert knowledge. Such explanations can be aligned with experimental data or expert-provided text or descriptors, rather than selected merely for statistical convenience, which may be physically incidental.

In this work, we propose {\em Geomo}, a framework that learns physically meaningful and actionable key motifs via a multimodal information bottleneck. Our goal is not to merely outperform black-box predictors on MAE, but to show that a motif discovered under chemical semantics can serve simultaneously as an inductive bias for prediction and as a physically consistent explanation.
In particular, {\em Geomo} encodes the 3D geometry of molecule--surface systems with an invariant geometric network that constructs hierarchical descriptors of increasing order and performs scalar message passing, and encodes expert textual descriptors with a tokenizer. A key motif identification module then learns a binary mask over geometric features to select atoms as key motifs. In addition to the traditional information bottleneck that only seeks minimal sufficient information relevant to prediction~\cite{tishby2015deeplearninginformationbottleneck}, we propose a multimodal bottleneck that adds a semantic anchor: the retained atoms must also be sufficient for the chemical context described by the experts. 
Because these descriptors name local chemical environments, this anchor steers motif selection toward local geometries that are both predictive and physically meaningful, rather than merely predictive. The resulting motifs are not post-hoc explanations; they serve as expert priors that guide the model's reasoning pathway, improving prediction accuracy and retaining competitive inference speed for high-throughput screening. The contributions are summarized as follows:

\begin{enumerate}
\item We introduce a expert textual knowledge-guided paradigm for structure--property modeling, in which expert descriptors of local chemical environments are used as semantic anchors for motif discovery, enabling physically meaningful structure--property relationships rather than merely predictive ones.


\item We propose a multimodal information bottleneck framework that formulates key motif identification as an information-theoretic optimization problem. It jointly optimizes a multimodal information bottleneck and a classical information bottleneck, derives tractable variational lower bounds, imposes a sparse Bernoulli prior on a learnable binary mask, and adopts a Gumbel-Softmax relaxation to enable end-to-end differentiable motif selection. This extracts compact geometric motifs that are maximally aligned with textual features while preserving their prediction sufficiency.

\item We develop a motif-centric modeling strategy in which the discovered motifs form the input pathway of the energy predictor, so that the predictor is trained through them rather than explained by them. This makes the motifs actionable in the sense of being used by the predictor and supporting molecule–surface system screening.


\item We curate a differential charge density dataset of molecule--surface systems and propose two metrics to quantify how physically meaningful the identified local geometries are, showing that the motifs align with electronic-structure quantities beyond what dedicated GNN explainers achieve.

\item For the representative task of catalytic configuration discovery in molecule--surface systems, {\em Geomo} achieves competitive energy prediction across diverse out-of-distribution splits, and its motifs quantitatively correlate with charge density maps, confirming that the model learns physically meaningful structure--property relationships.

\end{enumerate}

\section{Preliminaries}
\label{sec:related_work}

\subsection{Graph Neural Network Explanation}

Graph neural network explanation aims to reveal the graph components that contribute most to model predictions. Existing approaches can generally be categorized into instance-level and model-level explanations. Instance-level methods seek to identify important nodes, edges, or subgraphs for a specific input. GNNExplainer~\cite{ying2019gnnexplainer},
one of the earliest representative methods, learns continuous masks over graph structures to identify a compact subgraph that preserves the prediction of the original GNN. PGExplainer~\cite{luo2020pgexplainer} extends this paradigm to the inductive setting by parameterizing the explanation process with a neural network, allowing explanations to
generalize across graph instances. Subsequent studies further improve graph explanations through alternative subgraph selection and matching strategies. For example, MatchExplainer~\cite{pmlrv202wu23jExplaining} formulates explanation as a non-parametric subgraph matching problem. In parallel, attribution-based techniques such as GradCAM~\cite{selvaraju2017grad} estimate the importance of individual model components from gradient information.

Model-level approaches, in contrast, aim to characterize the global decision behavior of a GNN rather than explaining individual samples. For example, XGNN~\cite{yuan2020xgnn} generates representative graph patterns associated with target predictions. Although these methods have substantially advanced the interpretability of graph neural networks, most were originally developed for conventional graph representations and do not explicitly account for the geometric characteristics of three-dimensional atomic systems.





Despite their differences, these methods primarily follow a \emph{post-hoc} explanation paradigm: a predictive model is first trained, and an additional explanation procedure is subsequently used to identify the input components responsible for its output. The resulting explanations are therefore mainly optimized to reproduce or preserve the behavior of the underlying predictor. High explanation fidelity, however, does not necessarily imply that the identified structures correspond to the physical factors governing the target property.

\begin{figure*}[t!]
    \centering
   \includegraphics[width=0.999\textwidth]{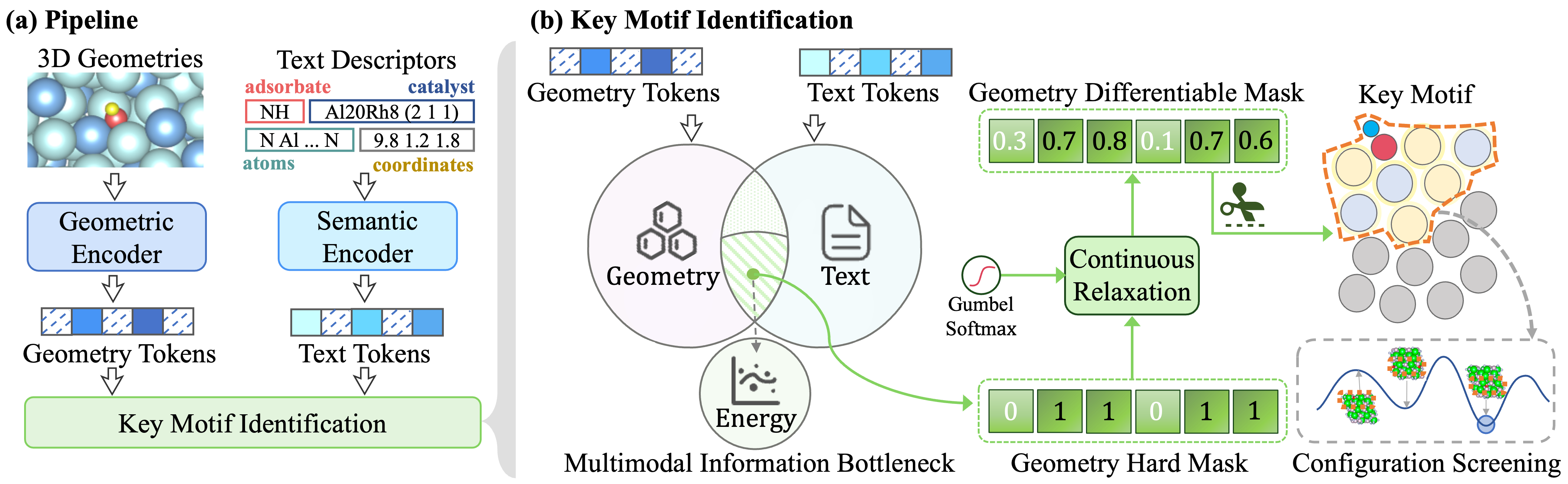}
    \caption{(a) {\em Geomo} encodes 3D geometries and textual descriptors using a geometric neural network and a tokenizer-based text encoder, respectively. A Key Motif Identification module then identifies key geometric motifs. (b) Key Motif Identification is designed based on the information bottleneck principle. It extracts compact geometric motifs that align with textual features while retaining sufficient information for accurate energy prediction by optimizing a learnable geometric mask. These motifs serve as both actionable explanations for adsorption behavior and inductive biases to improve relaxed energy prediction. Equipped with this capability, our method rapidly and accurately identifies the optimal configuration with the globally minimum adsorption energy.}
    \label{model}
\end{figure*}

\subsection{Typical Systems Studied: Catalytic Configurations}

Molecule–surface systems instantiate a broader class of problems where a global property is governed by a local geometric environment on an extended substrate. The proposed framework is motivated by this structure, and catalytic configurations serve as a concrete instantiation in our work.

Three concepts about the catalytic configurations are introduced: (i) the catalytic surface, the exposed surface of a solid catalyst that provides binding sites for adsorbates; (ii) the adsorbate, a molecule or atom that binds to the surface; and (iii) the adsorption configuration, the atomistic arrangement of the adsorbate on the surface, including the binding site, the adsorbate orientation, and the relaxed interatomic distances~\cite{kolluru2024adsorbdiff}. 
As illustrated by the 3D geometry in the upper-left corner of Fig. \ref{model}, the dark-blue and light-blue atoms form the catalytic surface, and the red and yellow atoms above represent the adsorbate molecule. Adsorbate binding to the surface gives rise to an adsorption configuration. The adsorption site, orientation, and arrangement of the adsorbate on the surface directly govern the catalytic performance.

Because different configurations yield different adsorption energies, we need to identify the optimal catalytic configuration, i.e., the configuration with globally minimized energy, as it represents the most stable binding state. Our goal is to predict the adsorption energy for a large number of candidate configurations and select the optimal one. This setting is a natural testbed for surrogate modeling: the space of possible configurations is combinatorially large, first-principles labels are expensive, and the target property is highly sensitive to local 3D structure.

\section{Methodology}

As shown in Figure \ref{model}(a), {\em Geomo} processes the 3D geometry $\mathcal{G}$ and the textual descriptors $\mathcal{S}$ using a geometric neural network and a text encoder, respectively.
For geometric representations $\mathcal{G}=(\mathcal{V},\mathcal{E})$, the node set $\mathcal{V}=\{v_1,\dots,v_N\}$ is characterized by a coordinate matrix $\mathbf{X}=[\mathbf{x}_1,\dots,\mathbf{x}_N]^T\in\mathbb{R}^{N\times 3}$. The edge set $\mathcal{E}$ is constructed under periodic boundary conditions: $\mathcal{E}=\{(i,j,\mathbf{k}) \mid \|\mathbf{x}_i-(\mathbf{x}_j+\mathbf{L}\mathbf{k})\|_2 \le R_{\text{cut}}\}$, where $\mathbf{L}\in\mathbb{R}^{3\times 3}$ is the lattice matrix, $\mathbf{k}$ is the cell offset vector, and $R_{\text{cut}}$ is the cutoff radius. By enumerating the offset vectors $\mathbf{k}$, the Euclidean distance between atom $i$ and other atoms $j$ is calculated. Candidate edges are generated when the distance is within $R_{\text{cut}}$, and the $K$ nearest neighbors among these candidates are retained.

We adopt the descriptors from Ock et al.~\cite{ock2023catalyst}, which summarize the adsorbate formula $\texttt{Adsorbate}$, adsorption site coordinates $\texttt{Position}$, surface element symbols $\texttt{Surface}$, Miller indices $\texttt{Miller}$, the primary atoms $\texttt{Primary}$, site types $\texttt{Site}$, and secondary atoms $\texttt{Secondary}$. 
The descriptors provide a compact chemical context for the adsorption environment.
All attributes are concatenated in a fixed order and then tokenized to form the token representation: $\mathcal{S} = \text{Tokenizer}(\texttt{Adsorbate} \oplus \texttt{Position} \oplus \texttt{Surface} \oplus \texttt{Miller} \oplus \texttt{Primary} \oplus \texttt{Site}\oplus \texttt{Secondary})$. These descriptors are naturally interpretable to experts and explicitly reflect that the local chemical environment governs adsorption behavior.

Given the representations $\mathcal{G}$ and $\mathcal{S}$, the geometric neural network and the text encoder produce geometry tokens and text tokens, respectively. Based on these tokens, as shown in Figure~\ref{model}(b), we introduce a key motif identification module that quantifies the cross-modal correlation between geometric and textual features, reflecting how strongly each geometric atom associates with textual expert knowledge and relates to adsorption behavior. This module implements a multimodal information bottleneck that extracts geometric features maximally aligned with textual knowledge and most informative for energy prediction, via a learnable geometric mask. The masked geometric atoms are then extracted as the key geometric motifs $\mathcal{G}_{\text{local}}$.
These key geometric motifs $\mathcal{G}_{\text{local}}$ provide actionable explanations: they serve as a physical prior for a parameterized model to predict the relaxed energy. 


\subsection{Geometric Encoder}
As adsorption energy is an invariant scalar, we design our model to operate entirely within an invariant framework. High-order geometric descriptors are constructed to capture local environments, and all message-passing updates are performed on scalar representations.
Specifically, we construct a hierarchy of invariant geometric descriptors from local atomic environments.
The local environment of each atom is represented by descriptors of increasing representation order. The two-order descriptor captures radial interactions,
\begin{equation}
\mathbf g_i^{(2)}
=
\frac{1}{|\mathcal N(i)|}
\sum_{j\in\mathcal N(i)}
\phi_2(d_{ij}),
\end{equation}
where $d_{ij}=\|\mathbf r_{ij}\|$, and $\phi_2$ consists of a radial basis expansion followed by a multilayer perceptron.
To encode angular geometry, we further construct a three-order descriptor,
\begin{equation}
\mathbf g_i^{(3)}=\frac{1}{|\mathcal A(i)|}\sum_{(j,k)\in\mathcal A(i)}\phi_3\left(d_{ij},d_{ik},\cos\theta_{jik}\right),
\end{equation}
where $\mathcal A(i)=\left\{(j,k)\mid j,k\in\mathcal N(i),~j\neq k \right\}$ and $\cos\theta_{jik}=\frac{\mathbf r_{ij}^{\top}\mathbf r_{ik}}{\|\mathbf r_{ij}\|\,\|\mathbf r_{ik}\|}.$
Rather than explicitly enumerating increasingly expensive higher-order tuples, descriptors of high order $s\ge4$ are recursively constructed by composing lower-order invariant descriptors,
\begin{equation}
\mathbf g_i^{(s)}=\phi_s\left(\mathbf g_i^{(2)},\mathbf g_i^{(3)},\ldots,\mathbf g_i^{(s-1)}\right),\qquad s\ge4,
\end{equation}
where $\phi_s$ denotes a shared multilayer perceptron.
This hierarchical composition progressively captures increasingly complex coordination motifs while avoiding the exponential complexity associated with explicit higher-order enumeration.
The complete local geometric representation is obtained by
\begin{equation}
\mathbf h_i^{(0)}=\phi_f\left(\bigoplus_{s=2}^{S=5}\mathbf g_i^{(s)}\right),
\end{equation}
where $\oplus$ denotes feature concatenation and $\phi_f$ projects descriptors from different orders into a common latent space.
The hierarchical descriptors are subsequently refined through scalar message passing,
\begin{equation}
\mathbf m_{ij}^{(t)}=\phi_m\left(\mathbf h_i^{(t)}\oplus \mathbf h_j^{(t)}\oplus\phi_e(d_{ij})\right),
\end{equation}
where $\phi_e$ encodes pairwise distances using radial basis functions.
Node representations are updated by Layer Normalization (LN)
\begin{equation}
\mathbf h_i^{(t+1)}=\mathrm{LN}\left(\mathbf h_i^{(t)}+\sum_{j\in\mathcal N(i)}\mathbf m_{ij}^{(t)}\right).
\end{equation}

Since all geometric descriptors are constructed exclusively from invariant quantities and message passing involves only scalar features, the learned representations remain invariant.

\subsection{Textual Descriptor Tokenizer}  
The textual descriptors $\mathcal{S}$ are processed by a text encoder based on the RoBERTa architecture~\cite{delobelle2020robbert}. Given initial representations $\mathbf{\mathbf{F}_S}^{(0)}=\mathcal{S}$, we use a stack of $M$ Transformer layers containing multi-head self-attention (MHA) and feed-forward networks (FFN) to encode the textual representations. The update rule for the textual token representations is:
\begin{equation}
\mathbf{\mathbf{F}_S}^{(l+1)} = \text{FFN}(\text{MHA}(\mathbf{\mathbf{F}_S}^{(l)})),
\end{equation}
where $\text{MHA}$ captures long-range dependencies among the chemical descriptors, and $\text{FFN}$ is responsible for the non-linear transformations. 

To align these token representations with the geometric features, the final semantic features $\mathbf{\mathbf{F}_S}$ are derived by projecting the token representations onto the target dimension $d$ and normalizing them:
\begin{equation}
\mathbf{F}_S = \text{Norm}(\text{Linear}(\mathbf{F}_S^{(M)})) \in \mathbb{R}^{L \times d},
\end{equation}
where $L$ is the number of tokens in the textual representation. This provides the critical expert knowledge necessary for identifying the key local regions of the adsorption system.

\subsection{Key Motif Identification}
Key motif identification extracts critical geometric structures based on two information-theoretic criteria: (i) relevance to textual semantic knowledge, and (ii) relevance to energy prediction. We formulate this as a joint information bottleneck problem whose bottleneck variable is the identified key geometric representation \(\mathbf{G}_{\mathrm{m}}\). A multimodal information bottleneck aligns \(\mathbf{G}_{\mathrm{m}}\) with textual semantics, while a classical information bottleneck preserves its predictive sufficiency for adsorption energy.

\textbf{Multimodal information bottleneck.}
Let
$\mathbf{H}=[\mathbf{F}_1;\dots;\mathbf{F}_{N_{\mathrm{atom}}}]\in\mathbb{R}^{N_{\mathrm{atom}}\times d}$
denote the atom representations produced by the geometric encoder, and let
$\mathbf{F}_{S,1},\dots,\mathbf{F}_{S,L}\in\mathbb{R}^{d}$
denote the $L$ token representations produced by the text encoder. We summarize the textual expert knowledge into a pooled semantic vector
\begin{equation}
\bar{\mathbf{F}}_S=\frac{1}{L}\sum_{l=1}^{L}\mathbf{F}_{S,l}\in\mathbb{R}^{d}.
\end{equation}
We introduce a binary mask $\mathbf{M}\in\{0,1\}^{N_{\mathrm{atom}}}$ and define the key geometirc representation $\mathbf{G}_{\mathrm{m}}=\mathbf{H}\odot\mathbf{M}$, where $\mathbf{G}_{\mathrm{m},i}=M_i\mathbf{F}_i$. The criterion is that $\mathbf{G}_{\mathrm{m}}$ should preserve the chemical context described by the textual expert knowledge. We encode this criterion as
\begin{equation}
\max_{\theta,\phi}\; I(\mathbf{G}_{\mathrm{m}},\bar{\mathbf{F}}_S)-\beta I(\mathbf{H},\mathbf{G}_{\mathrm{m}}),
\end{equation}
where the first term encourages the retained atoms to preserve textual semantics, and the second term compresses the selection to be sparse and non-redundant.

\emph{Compression term.}
We first bound $I(\mathbf{H},\mathbf{G}_{\mathrm{m}})$. Since $\mathbf{G}_{\mathrm{m}}$ is a deterministic function of $\mathbf{M}$ given $\mathbf{H}$, the inequality gives $I(\mathbf{H},\mathbf{G}_{\mathrm{m}})\le I(\mathbf{H},\mathbf{M})$.
We then expand the mutual information between $\mathbf{H}$ and $\mathbf{M}$:
\begin{equation}
\begin{aligned}
I(\mathbf{H},\mathbf{M})
&= \mathbb{E}_{p(\mathbf{H},\mathbf{M})}
\left[
\log\frac{p(\mathbf{M}\mid\mathbf{H})}{p(\mathbf{M})}
\right] \\
&= \mathbb{E}_{\mathbf{H}\sim p(\mathbf{H})}
\mathrm{KL}\!\left(
p(\mathbf{M}\mid\mathbf{H})\,\|\,p(\mathbf{M})
\right).
\end{aligned}
\end{equation}
Both the conditional posterior $p(\mathbf{M}\mid\mathbf{H})$ and the marginal $p(\mathbf{M})$ are intractable. To obtain a tractable bound, we introduce a prior $r(\mathbf{M})$ over the mask and use the non-negativity of $\mathrm{KL}(p(\mathbf{M})\,\|\,r(\mathbf{M}))$ to write
\begin{equation}
\begin{aligned}
I(\mathbf{H},\mathbf{M})
&= \mathbb{E}_{\mathbf{H}\sim p(\mathbf{H})}
\mathrm{KL}\!\left(p(\mathbf{M}\mid\mathbf{H})\,\|\,r(\mathbf{M})\right) \\
&\quad - \mathrm{KL}\!\left(p(\mathbf{M})\,\|\,r(\mathbf{M})\right) \\
&\le \mathbb{E}_{\mathbf{H}\sim p(\mathbf{H})}
\mathrm{KL}\!\left(p(\mathbf{M}\mid\mathbf{H})\,\|\,r(\mathbf{M})\right).
\end{aligned}
\end{equation}
The remaining conditional posterior $p(\mathbf{M}\mid\mathbf{H})$ is still unknown, so we parameterize it with a neural network
$q_\theta(\mathbf{M}\mid\mathbf{H})$:
\begin{equation}
q_\theta(\mathbf{M}\mid\mathbf{H})
=
\prod_{i=1}^{N_{\mathrm{atom}}}
\mathrm{Bernoulli}(M_i;\,p_i),
\qquad
p_i=\sigma\!\left(f_\theta(\mathbf{F}_i)\right),
\end{equation}
where $f_\theta$ is a two-layer MLP that outputs a scalar mask logit for each atom. We choose the sparse prior
\begin{equation}
r(\mathbf{M})
=
\prod_{i=1}^{N_{\mathrm{atom}}}
\mathrm{Bernoulli}(M_i;\,0.1),
\end{equation}
which selects each atom with probability $0.1$, reflecting the belief that only a small fraction of atoms form the key motif. Since both $q_\theta(\mathbf{M}\mid\mathbf{H})$ and $r(\mathbf{M})$ are independent Bernoulli, the KL term admits the closed form
\begin{equation}
\mathcal{D}_{\mathrm{KL}}
=
\sum_{i=1}^{N_{\mathrm{atom}}}
\left[
p_i\log\frac{p_i}{0.1}
+
(1-p_i)\log\frac{1-p_i}{0.9}
\right].
\end{equation}
This term penalizes masks whose selection probabilities deviate from the sparse prior and thus drives the model toward compact atom selections.
Considering all samples in the training set, we define the KL loss as the average Bernoulli KL divergence between the mask posterior \(q_\theta(\mathbf{M}\mid\mathbf{H})\) and the sparse prior \(r(\mathbf{M})\):
\begin{equation}
\mathcal{L}_{\mathrm{KL}}
=
\frac{1}{N}
\sum_{n=1}^{N}
\sum_{i=1}^{N_{\mathrm{atom}}^{(n)}}
\left[
p_i^{(n)}\log\frac{p_i^{(n)}}{0.1}
+
(1-p_i^{(n)})\log\frac{1-p_i^{(n)}}{0.9}
\right],
\label{eq:kl_loss}
\end{equation}
where \(p_i^{(n)}=\sigma\!\left(f_\theta(\mathbf{F}_i^{(n)})\right)\) is the selection probability of atom \(i\) in sample \(n\), \(N\) is the number of training samples, and \(N_{\mathrm{atom}}^{(n)}\) is the number of atoms in sample \(n\).

\emph{Semantic alignment term.}
We now bound $I(\mathbf{G}_{\mathrm{m}},\bar{\mathbf{F}}_S)$. Using the entropy decomposition, we write
\begin{equation}
I(\mathbf{G}_{\mathrm{m}},\bar{\mathbf{F}}_S)
=
H(\bar{\mathbf{F}}_S)-H(\bar{\mathbf{F}}_S\mid\mathbf{G}_{\mathrm{m}}),
\end{equation}
where $H(\bar{\mathbf{F}}_S)$ is the differential entropy of the pooled textual representation and is determined by the data distribution alone. The conditional entropy
\begin{equation}
H(\bar{\mathbf{F}}_S\mid\mathbf{G}_{\mathrm{m}})
=
-\mathbb{E}_{p(\mathbf{G}_{\mathrm{m}},\bar{\mathbf{F}}_S)}
\big[\log p(\bar{\mathbf{F}}_S\mid\mathbf{G}_{\mathrm{m}})\big]
\end{equation}
depends on the intractable conditional likelihood $p(\bar{\mathbf{F}}_S\mid\mathbf{G}_{\mathrm{m}})$. We therefore introduce a variational distribution $q_\phi(\bar{\mathbf{F}}_S\mid\mathbf{G}_{\mathrm{m}})$ with parameters $\phi$. This distribution maps from the masked geometric representation $\mathbf{G}_{\mathrm{m}}$ to the textual semantic vector $\bar{\mathbf{F}}_S$; intuitively, $q_\phi$ acts as a decoder that reconstructs the textual semantics from the retained atoms. Since
\begin{equation}
\mathrm{KL}\!\left(p(\bar{\mathbf{F}}_S\mid\mathbf{G}_{\mathrm{m}})\,\|\,q_\phi(\bar{\mathbf{F}}_S\mid\mathbf{G}_{\mathrm{m}})\right)\ge 0,
\end{equation}
we have
\begin{equation}
H(\bar{\mathbf{F}}_S\mid\mathbf{G}_{\mathrm{m}})
\le
-\mathbb{E}_{p(\mathbf{G}_{\mathrm{m}},\bar{\mathbf{F}}_S)}
\big[\log q_\phi(\bar{\mathbf{F}}_S\mid\mathbf{G}_{\mathrm{m}})\big].
\end{equation}
Substituting this bound into the entropy decomposition yields
\begin{equation}
I(\mathbf{G}_{\mathrm{m}},\bar{\mathbf{F}}_S)
\ge
H(\bar{\mathbf{F}}_S)
+
\mathbb{E}_{\mathbf{H},\mathbf{M}}
\big[\log q_\phi(\bar{\mathbf{F}}_S\mid\mathbf{G}_{\mathrm{m}})\big],
\label{eq:sem_lb}
\end{equation}
where $\mathbb{E}_{\mathbf{H},\mathbf{M}}$ abbreviates $\mathbb{E}_{\mathbf{H}\sim p(\mathbf{H})}\mathbb{E}_{\mathbf{M}\sim q_\theta(\mathbf{M}\mid\mathbf{H})}$.
Since $H(\bar{\mathbf{F}}_S)$ is determined by the data distribution and is constant, it can be omitted from the training objective.

We model $q_\phi(\bar{\mathbf{F}}_S\mid\mathbf{G}_{\mathrm{m}})$ as a Gaussian distribution over the semantic vector $\bar{\mathbf{F}}_S$, conditioned on the masked representation $\mathbf{G}_{\mathrm{m}}$, with identity covariance,
\begin{equation}
q_\phi(\bar{\mathbf{F}}_S\mid\mathbf{G}_{\mathrm{m}})
=
\mathcal{N}\!\left(\bar{\mathbf{F}}_S;\,\mu_\phi(\mathbf{G}_{\mathrm{m}}),\,\mathbf{I}_d\right),
\end{equation}
where $\mu_\phi(\mathbf{G}_{\mathrm{m}})\in\mathbb{R}^{d}$ is the mean produced by the decoder network. Note that $q_\phi(\cdot\mid\mathbf{G}_{\mathrm{m}})$ is a distribution rather than a point estimate: the decoder network only needs to output the mean $\mu_\phi(\mathbf{G}_{\mathrm{m}})$, because the covariance is fixed to $\mathbf{I}_d$. The actual prediction is therefore the Gaussian distribution, and $\mu_\phi(\mathbf{G}_{\mathrm{m}})$ is its center.
This Gaussian choice serves two purposes. First, it yields a closed-form likelihood, making the variational lower bound tractable. Second, with identity covariance, the Gaussian negative log-likelihood reduces to a squared reconstruction error,
\begin{equation}
-\log q_\phi(\bar{\mathbf{F}}_S\mid\mathbf{G}_{\mathrm{m}})
=
\frac12\|\bar{\mathbf{F}}_S-\mu_\phi(\mathbf{G}_{\mathrm{m}})\|_2^2+\mathrm{const}.
\label{eq:sem_gaussian}
\end{equation}
During training, we evaluate the likelihood of the observed $\bar{\mathbf{F}}_S$ under this distribution; maximizing the likelihood is therefore equivalent to pushing $\mu_\phi(\mathbf{G}_{\mathrm{m}})$ toward $\bar{\mathbf{F}}_S$, i.e., minimizing the expected reconstruction error over the mask posterior. Fixing the covariance avoids an additional variance parameter and prevents the model from reducing the negative log-likelihood by inflating the variance instead of improving the mean.

The decoder network produces $\mu_\phi(\mathbf{G}_{\mathrm{m}})$ in two steps. It first pools the retained atoms by masked mean pooling,
\begin{equation}
\mathrm{Pool}(\mathbf{G}_{\mathrm{m}})
=
\frac{\sum_{i=1}^{N_{\mathrm{atom}}}\mathbf{G}_{\mathrm{m},i}}
{\max\!\left(\sum_{i=1}^{N_{\mathrm{atom}}}M_i,\,\varepsilon\right)},
\end{equation}
where $\varepsilon>0$ is a small numerical constant (e.g., $10^{-6}$) that prevents division by zero when the mask selects no atom. It then maps the pooled vector to the semantic space,
\begin{equation}
\mu_\phi(\mathbf{G}_{\mathrm{m}})
=
\mathrm{LN}\!\left(
\mathbf{W}_2\,\mathrm{SiLU}(\mathbf{W}_1\,\mathrm{Pool}(\mathbf{G}_{\mathrm{m}})+\mathbf{b}_1)+\mathbf{b}_2
\right).
\end{equation}
Here $\mathrm{Pool}$ and $\mu_\phi$ are deterministic mappings, and the parameters $\{\mathbf{W}_1,\mathbf{b}_1,\mathbf{W}_2,\mathbf{b}_2\}\subset\phi$ are shared across samples. Because masked mean pooling only aggregates retained atoms, the decoder can reconstruct $\bar{\mathbf{F}}_S$ only if the mask preserves atoms that carry the textual semantics.

\textbf{Energy information bottleneck.}
The second criterion is that the masked geometric representation should remain sufficient for energy prediction. We encode this criterion as a second information bottleneck,
\begin{equation}
\max_{\theta,\psi}\;
I(\mathbf{G}_{\mathrm{m}},\hat{E})-\beta I(\mathbf{H},\mathbf{G}_{\mathrm{m}}),
\end{equation}
with the same mask posterior \(q_\theta(\mathbf{M}\mid\mathbf{H})\) and prior \(r(\mathbf{M})\), so the compression term admits the same KL regularization derived above. The new term is the predictive term \(I(\mathbf{G}_{\mathrm{m}},\hat{E})\). To lower-bound it, we introduce a variational likelihood \(q_\psi(\hat{E}\mid\mathbf{G}_{\mathrm{m}})\) that predicts the adsorption energy from the masked representation:
\begin{equation}
q_\psi(\hat{E}\mid\mathbf{G}_{\mathrm{m}})
=
\mathcal{N}\!\left(\hat{E};\,\hat{y}(\mathbf{G}_{\mathrm{m}}),\,\sigma^2\right),
\end{equation}
where the mean is an atom-level average over the retained atoms,
\begin{equation}
\hat{y}(\mathbf{G}_{\mathrm{m}})
=
\frac{\sum_{i=1}^{N_{\mathrm{atom}}}M_i\,\mathrm{FFN}(\mathbf{F}_i)}
{\max(\sum_{i=1}^{N_{\mathrm{atom}}}M_i,\,\varepsilon)},
\end{equation}
and \(\mathrm{FFN}\) maps each atom representation to a scalar energy contribution. The denominator normalizes by the number of retained atoms, so \(\hat{y}(\mathbf{G}_{\mathrm{m}})\) remains comparable across different mask realizations. Setting \(\sigma=1\), the Gaussian negative log-likelihood reduces to a squared error,
\begin{equation}
-\log q_\psi(\hat{E}\mid\mathbf{G}_{\mathrm{m}})
=
\frac12\big(\hat{E}-\hat{y}(\mathbf{G}_{\mathrm{m}})\big)^2+\mathrm{const},
\label{eq:energy_lb}
\end{equation}
and the corresponding variational lower bound is
\begin{equation}
I(\mathbf{G}_{\mathrm{m}},\hat{E})
\ge
\mathbb{E}_{\mathbf{H}\sim p(\mathbf{H})}
\mathbb{E}_{\mathbf{M}\sim q_\theta(\mathbf{M}\mid\mathbf{H})}
\big[\log q_\psi(\hat{E}\mid\mathbf{G}_{\mathrm{m}})\big].
\end{equation}
Thus, maximizing the predictive term is equivalent to minimizing the expected squared energy error over the mask posterior.

\textbf{Continuous relaxation.}
The two bounds above require sampling \(\mathbf{M}\sim q_\theta(\mathbf{M}\mid\mathbf{H})\), which is non-differentiable because \(\mathbf{M}\) is discrete, and hence \(\mathbf{G}_{\mathrm{m}}=\mathbf{H}\odot\mathbf{M}\) is also non-differentiable. We therefore replace the discrete sample with a continuous surrogate using the Gumbel-Sigmoid variant of the Gumbel-Softmax trick. Let \(a_i=f_\theta(\mathbf{F}_i)\) be the mask logit, so that \(p_i=\sigma(a_i)\). With Gumbel noise \(g_i=-\log(-\log e_i)\), \(e_i\sim\mathcal{U}(0,1)\), the relaxed mask entry is
\begin{equation}
c_i
=
\sigma\!\left(\frac{a_i+g_i}{\tau}\right),
\end{equation}
where \(\tau\) is the temperature parameter. A large \(\tau\) yields smooth selections, while a small \(\tau\) approximates hard \(0/1\) selection; \(\tau\) is annealed from \(1.0\) to \(0.1\) during training. To stabilize selection and avoid mask fluctuations, we perform \(K\) independent Gumbel samplings and take the element-wise maximum,
\begin{equation}
M_i^{*}
=
\max_{k=1,\dots,K}c_i^{(k)},
\qquad
c_i^{(k)}
=
\sigma\!\left(\frac{a_i+g_i^{(k)}}{\tau}\right),
\end{equation}
with \(K=5\). The relaxed mask vector \(\mathbf{M}^*\in(0,1)^{N_{\mathrm{atom}}}\) defines a differentiable surrogate for the bottleneck,
\begin{equation}
\mathbf{G}_{\mathrm{m}}^{*}=\mathbf{H}\odot\mathbf{M}^{*},
\qquad
\mathbf{G}_{\mathrm{m},i}^{*}=M_i^{*}\mathbf{F}_i,
\end{equation}
which is used to instantiate the expectations above via Monte Carlo. The element-wise maximum over \(K\) samples sharpens the mask but does not binarize it. Binarization is performed only at inference, where the continuous mask is thresholded at \(0.5\) via \(M_i=\mathbb{I}(M_i^*>0.5)\).

\textbf{Overall training objective.}
Both information bottleneck objectives are maximization problems, so the corresponding training losses are their negative variational lower bounds, with the expectation over $\mathbf{M}\sim q_\theta(\mathbf{M}\mid\mathbf{H})$ approximated by Monte Carlo using the relaxed mask $\mathbf{G}_{\mathrm{m}}^{*}$. Let $N$ be the number of training samples, $N_{\mathrm{atom}}^{(n)}$ the number of atoms in sample $n$, $\mathbf{G}_{\mathrm{m}}^{*(n)}$ the relaxed masked representation, $\bar{\mathbf{F}}_S^{(n)}$ the pooled semantic vector, and $E^{(n)}$ the ground-truth relaxed adsorption energy of sample $n$.

The semantic reconstruction loss is the negative of the variational lower bound in Eq.~\eqref{eq:sem_lb}, instantiated with the Gaussian decoder in Eq.~\eqref{eq:sem_gaussian}:
\begin{equation}
\mathcal{L}_{\mathrm{rec}}
=
\frac{1}{N}
\sum_{n=1}^{N}
\frac12
\left\|
\bar{\mathbf{F}}_S^{(n)}
-
\mu_\phi\!\left(\mathbf{G}_{\mathrm{m}}^{*(n)}\right)
\right\|_2^2 .
\end{equation}
The energy loss is the negative of the variational lower bound in Eq.~\eqref{eq:energy_lb},
\begin{equation}
\mathcal{L}_{\mathrm{energy}}
=
\frac{1}{N}
\sum_{n=1}^{N}
\frac12
\left(
E^{(n)}
-
\hat{y}\!\left(\mathbf{G}_{\mathrm{m}}^{*(n)}\right)
\right)^2 .
\end{equation}
The KL loss is the average Bernoulli KL divergence already defined in Eq.~\eqref{eq:kl_loss}; since it depends only on the selection probabilities $p_i^{(n)}$ rather than on the sampled relaxed mask, it is computed exactly without Monte Carlo averaging.

Accordingly, the two bottleneck losses are
\begin{equation}
\mathcal{L}_{\mathrm{MIB}}
=
\mathcal{L}_{\mathrm{rec}}
+
\beta\mathcal{L}_{\mathrm{KL}},
\qquad
\mathcal{L}_{\mathrm{YIB}}
=
\mathcal{L}_{\mathrm{energy}}
+
\beta\mathcal{L}_{\mathrm{KL}},
\end{equation}
and the overall objective is
\begin{equation}
\mathcal{L}_{\mathrm{total}}
=
\mathcal{L}_{\mathrm{MIB}}
+
\mathcal{L}_{\mathrm{YIB}}.
\end{equation}
This objective jointly optimizes the mask generator, the semantic reconstruction head, and the energy predictor, yielding compact geometric motifs that are semantically aligned with textual expert knowledge and sufficient for accurate adsorption-energy prediction.

\section{Results and Discussion}

\subsection{Benchmark Datasets}


The Open Catalyst dataset~\cite{chanussot2021open} represents a milestone in the catalysis community, providing over 1.2 million DFT relaxations across 55 elements and 82 adsorbates. By offering a vast collection of nitrogen, carbon, and oxygen chemistries on diverse inorganic interfaces, the Open Catalyst database addresses the computational bottleneck of DFT in discovering catalysts for solar fuels synthesis and renewable energy storage. 

\begin{figure*}[htbp]
    \centering  \includegraphics[width=1.0\textwidth]{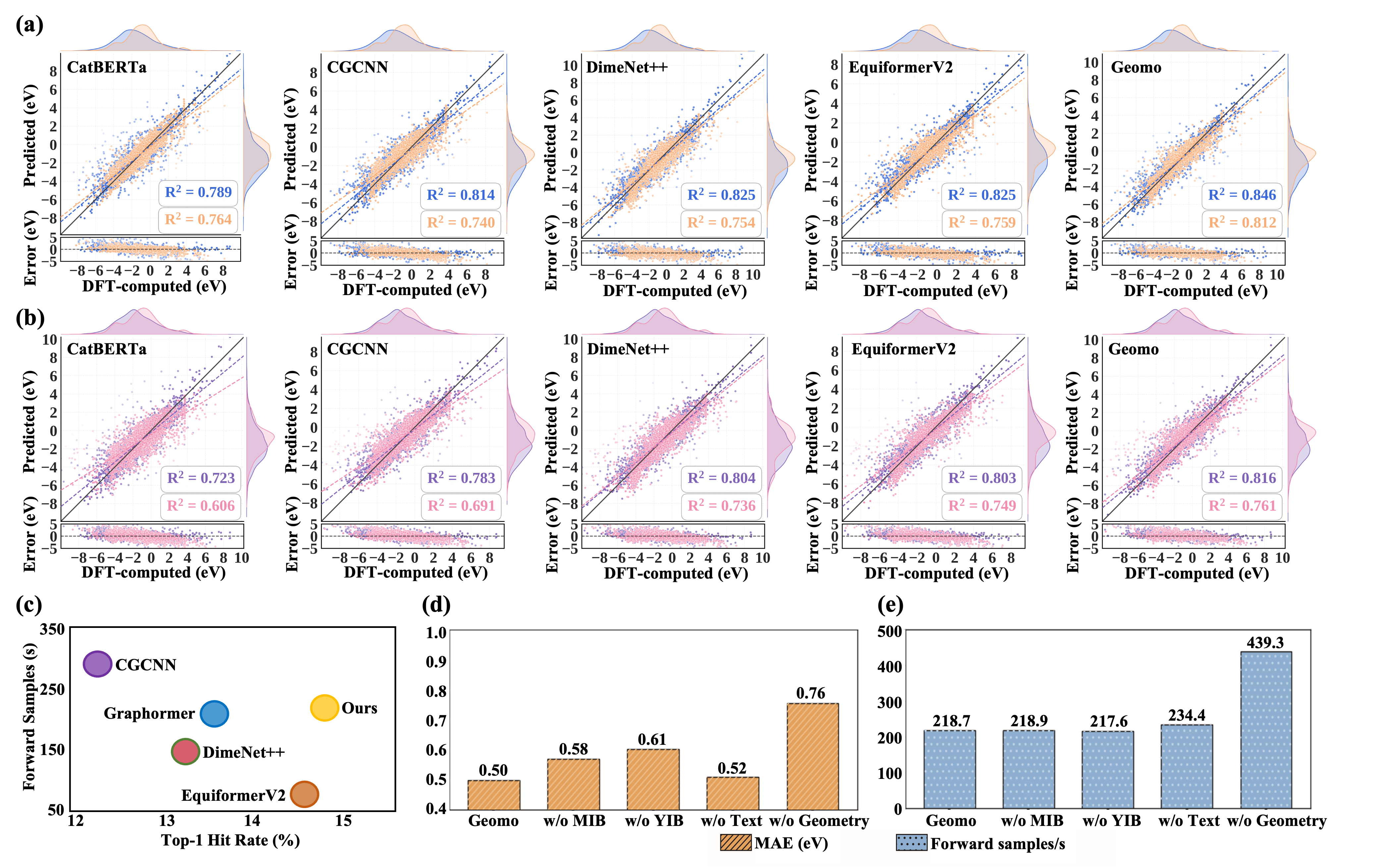}
    \caption{(a) Coefficient of determination $R^2$ for energy prediction on the 460k setting. Yellow and blue dots correspond to OOD$_{\text{ads}}$ and OOD$_{\text{both}}$, respectively. (b) Coefficient of determination $R^2$ on the 460k setting. Pink and purple dots denote in-domain and OOD$_{\text{cat}}$ data, respectively. (c) Top-1 hit rate and inference speed using different graph encoders. 
    (d) and (e) Ablation results averaged over all data splits (ID, OOD$_{\text{ads}}$, OOD$_{\text{cat}}$, OOD$_{\text{both}}$) and data scales (10k, 100k, 460k), in terms of MAE and inference speed.    }
    \label{R2460}
\end{figure*}

\textbf{OC20 IS2RE.} In this work, we focus on the Initial Structure to Relaxed Energy (IS2RE) task in OC20. Given an unrelaxed initial adsorption structure, the goal is to predict the final relaxed adsorption energy without carrying out DFT relaxation. We evaluate the model under three training settings: Train-10k, Train-100k, and the full-scale 460k setting. These settings allow us to examine model behavior in both low-data and large-scale regimes. 
Since the DFT relaxation trajectories for the test data set are not publicly accessible, our evaluation and comparison of model performance are centered on the validation data set.
The database provides four official validation splits to systematically evaluate model generalization: The ID split contains adsorbates and catalysts seen during training but uses new geometries. The {OOD}$_{\mathrm{ads}}$ split contains unseen adsorbates on known catalysts, while the {OOD}$_{\mathrm{cat}}$ split contains known adsorbates on unseen catalyst compositions. The {OOD}$_{\mathrm{both}}$ split contains both unseen adsorbates and unseen catalyst materials.

\textbf{OC20Dense.} Beyond energy prediction on OC20 IS2RE, we also evaluate high-throughput adsorption configuration screening on OC20Dense. Unlike standard IS2RE, where each adsorption structure is treated as an independent regression example, OC20Dense provides multiple candidate adsorption configurations for the same adsorbate--catalyst system. In this benchmark, candidate configurations are ranked by predicted relaxed adsorption energy, and the lowest-energy configuration selected by the model is compared with the DFT global minimum. This setting tests whether a model can identify the most stable adsorption configuration from a dense candidate pool, a central requirement in practical catalyst screening.

OC20Dense complements the OC20 IS2RE regression benchmark by moving the evaluation from sample-level energy prediction to system-level configuration selection. It also measures ranking quality and the energy penalty caused by selecting a suboptimal adsorption configuration. Together, these benchmarks assess whether relaxation-free adsorption energy predictors can support both accurate energy estimation and efficient high-throughput configuration screening.

\textbf{Baselines.} We compare \emph{Geomo} with three categories of baselines: 1) Textual representation-based methods, such as Graphormer~\cite{ying2021transformers}, CatBERTa~\cite{ock2023catalyst}. 2) For structure-based methods, we include CGCNN~\cite{xie2018crystal}, DimeNet++~\cite{gasteiger2020directional}, EquiformerV2~\cite{liao2023equiformerv2}, and Graphormer~\cite{ying2021transformers}. 3) We also include two hybrid or multimodal baselines, GAP~\cite{ock2024multimodal} and AdsMT~\cite{chen2025multi}. Among these baselines, CGCNN and DimeNet++ use officially available IS2RE results or checkpoints, while Graphormer, CatBERTa, GAP, and EquiformerV2 were trained using publicly available source code, as pre-trained weights for the IS2RE task have not been released. Detailed hyperparameter settings and parameter sizes are provided in the Supplemental Materials.


For explanation evaluation, we compare {\em Geomo} with the GNNExplainer~\cite{ying2019gnnexplainer} and MatchExplainer~\cite{pmlrv202wu23jExplaining}. Also, to endow existing black-box methods with explainability, we equip CGCNN, DimeNet++, EquiformerV2, and CatBERTa with the classical explainability technique GradCAM~\cite{selvaraju2017grad}.

\subsection{Motif-Based Relaxed Energy Prediction}

\textbf{Results and Analysis.} The motif-based relaxed energy prediction task focuses on the IS2RE task, which aims to predict the final relaxed energy directly from an unrelaxed initial state. Relaxed adsorption energies are essential descriptors for catalyst activity and selectivity in microkinetic modeling, and bypassing the costly iterative relaxation process.

\begin{table*}[t] 
\centering
\caption{MAE for energy prediction across various models. The best result is \textbf{bold}. All models are evaluated on the same validation splits; training configurations follow the official or publicly released settings. }
\label{mae}
\setlength{\tabcolsep}{10pt} 
\resizebox{0.8\textwidth}{!}{
    \small
    \begin{tabular}{ll cccc}
    \toprule
    \multicolumn{2}{c}{} & \multicolumn{4}{c}{\textbf{MAE [eV] $\downarrow$}} \\
    \cmidrule(lr){3-6}
    Data Size & Model & ID & OOD$_{\text{ads}}$ & OOD$_{\text{cat}}$ & OOD$_{\text{both}}$ \\
    \midrule
    10k & Graphormer~\cite{ying2021transformers} & 0.965$_{\pm0.005}$ & 0.994$_{\pm0.001}$ & 0.987$_{\pm0.012}$ & 1.018$_{\pm0.003}$ \\
    10k & CatBERTa~\cite{ock2023catalyst} & 0.935$_{\pm0.013}$ & 0.951$_{\pm0.004}$ & 0.908$_{\pm0.010}$ & 0.886$_{\pm0.009}$ \\
    10k & GAP~\cite{ock2024multimodal} & 0.834$_{\pm0.006}$ & 0.862$_{\pm0.006}$ & 0.831$_{\pm0.011}$ & 0.838$_{\pm0.015}$ \\
    10k & AdsMT~\cite{chen2025multi} & 0.838$_{\pm0.009}$ & 0.871$_{\pm0.007}$ & 0.796$_{\pm0.003}$ & 0.779$_{\pm0.006}$ \\
    10k & CGCNN~\cite{xie2018crystal} & 1.033$_{\pm0.022}$ & 1.005$_{\pm0.011}$ & 0.991$_{\pm0.007}$ & 0.881$_{\pm0.002}$ \\
    10k & DimeNet++~\cite{gasteiger2020directional} & 0.917$_{\pm0.018}$ & 0.972$_{\pm0.016}$ & 0.854$_{\pm0.009}$ & 0.872$_{\pm0.006}$ \\
    10k & EquiformerV2~\cite{liao2023equiformerv2} & 0.867$_{\pm0.015}$ & 0.906$_{\pm0.014}$ & 0.822$_{\pm0.012}$ & 0.802$_{\pm0.007}$ \\
    \rowcolor[gray]{0.9} 10k & {\em Geomo} & \textbf{0.758$_{\pm0.011}$} & \textbf{0.751$_{\pm0.008}$} & \textbf{0.711$_{\pm0.009}$} & \textbf{0.697$_{\pm0.006}$} \\
    \midrule
    100k & Graphormer~\cite{ying2021transformers} & 0.619$_{\pm0.009}$ & 0.812$_{\pm0.014}$ & 0.653$_{\pm0.005}$ & 0.745$_{\pm0.013}$ \\
    100k & CatBERTa~\cite{ock2023catalyst} & 0.779$_{\pm0.008}$ & 1.003$_{\pm0.004}$ & 0.749$_{\pm0.018}$ & 0.942$_{\pm0.007}$ \\
    100k & GAP~\cite{ock2024multimodal} & 0.696$_{\pm0.007}$ & 0.886$_{\pm0.012}$ & 0.647$_{\pm0.010}$ & 0.889$_{\pm0.014}$ \\
    100k & AdsMT~\cite{chen2025multi} & 0.689$_{\pm0.006}$ & 0.782$_{\pm0.011}$ & 0.664$_{\pm0.009}$ & 0.715$_{\pm0.008}$ \\
    100k & CGCNN~\cite{xie2018crystal} & 0.702$_{\pm0.019}$ & 0.869$_{\pm0.013}$ & 0.657$_{\pm0.002}$ & 0.760$_{\pm0.004}$ \\
    100k & DimeNet++~\cite{gasteiger2020directional} & 0.665$_{\pm0.015}$ & 0.822$_{\pm0.009}$ & 0.627$_{\pm0.008}$ & 0.748$_{\pm0.003}$ \\
    100k & EquiformerV2~\cite{liao2023equiformerv2} & 0.684$_{\pm0.010}$ & 0.762$_{\pm0.013}$ & 0.644$_{\pm0.006}$ & 0.675$_{\pm0.004}$ \\
    \rowcolor[gray]{0.9} 100k & {\em Geomo} & \textbf{0.598$_{\pm0.006}$} & \textbf{0.705$_{\pm0.008}$} & \textbf{0.589$_{\pm0.007}$} & \textbf{0.627$_{\pm0.009}$} \\
    \midrule
    460k & Graphormer~\cite{ying2021transformers} & 0.526$_{\pm0.004}$ & 0.704$_{\pm0.005}$ & 0.589$_{\pm0.009}$ & 0.627$_{\pm0.007}$ \\
    460k & CatBERTa~\cite{ock2023catalyst} & 0.668$_{\pm0.007}$ & 0.873$_{\pm0.013}$ & 0.637$_{\pm0.014}$ & 0.814$_{\pm0.005}$ \\
    460k & GAP~\cite{ock2024multimodal} & 0.613$_{\pm0.010}$ & 0.778$_{\pm0.009}$ & 0.574$_{\pm0.009}$ & 0.735$_{\pm0.008}$ \\
    460k & AdsMT~\cite{chen2025multi} & 0.472$_{\pm0.005}$ & 0.689$_{\pm0.007}$ & 0.505$_{\pm0.009}$ & 0.589$_{\pm0.003}$ \\
    460k & CGCNN~\cite{xie2018crystal} & 0.609$_{\pm0.006}$ & 0.822$_{\pm0.012}$ & 0.618$_{\pm0.005}$ & 0.756$_{\pm0.010}$ \\
    460k & DimeNet++~\cite{gasteiger2020directional} & 0.557$_{\pm0.009}$ & 0.731$_{\pm0.006}$ & 0.569$_{\pm0.010}$ & 0.656$_{\pm0.005}$ \\
    460k & EquiformerV2~\cite{liao2023equiformerv2} & 0.463$_{\pm0.006}$ & 0.582$_{\pm0.008}$ & 0.495$_{\pm0.005}$ & 0.557$_{\pm0.007}$ \\
    \rowcolor[gray]{0.9} 460k & {\em Geomo} & \textbf{0.432$_{\pm0.010}$} & \textbf{0.561$_{\pm0.007}$} & \textbf{0.462$_{\pm0.004}$} & \textbf{0.540$_{\pm0.006}$} \\
    \bottomrule
    \end{tabular}
}
\end{table*}

\begin{table*}[htbp]
\centering
\caption{Adsorption configuration screening results on the OC20Dense benchmark. The best result in each column is shown in \textbf{bold}.}
\label{oc20dense_screening}
\setlength{\tabcolsep}{6pt}
\resizebox{0.8\textwidth}{!}{
\small
\begin{tabular}{l c c c c}
\toprule
\multicolumn{1}{c}{\textbf{Model}} & \multicolumn{3}{c}{\textbf{Screening Metrics}} & \textbf{Efficiency} \\
\cmidrule(lr){2-4} \cmidrule(lr){5-5}
& Top-1 Hit (\%) $\uparrow$ & Top-3 Hit (\%) $\uparrow$ & Median Gap (eV) $\downarrow$ & Forward samples/s $\uparrow$ \\
\midrule
CatBERTa~\cite{ock2023catalyst} & 11.7$_{\pm1.6}$ & 34.2$_{\pm1.8}$ & 0.24$_{\pm0.06}$ & \textbf{439.30$_{\pm47.67}$} \\
GAP~\cite{ock2024multimodal} & 12.5$_{\pm1.2}$ & 34.9$_{\pm1.4}$ & 0.20$_{\pm0.04}$ & 397.31$_{\pm67.18}$ \\
CGCNN~\cite{xie2018crystal} & 12.0$_{\pm1.5}$ & 35.5$_{\pm2.0}$ & 0.22$_{\pm0.08}$ & 323.00$_{\pm31.74}$ \\
DimeNet++~\cite{gasteiger2020directional} & 12.8$_{\pm1.4}$ & 35.8$_{\pm1.3}$ & 0.18$_{\pm0.04}$ & 157.11$_{\pm18.11}$ \\
Graphormer~\cite{ying2021transformers} & 13.1$_{\pm0.9}$ & 36.3$_{\pm1.5}$ & 0.16$_{\pm0.04}$ & 213.53$_{\pm18.39}$ \\
EquiformerV2~\cite{liao2023equiformerv2} & 14.2$_{\pm1.2}$ & 36.6$_{\pm1.2}$ & 0.14$_{\pm0.02}$ & 89.18$_{\pm15.30}$ \\
AdsMT~\cite{chen2025multi} & 12.2$_{\pm1.3}$ & 35.2$_{\pm1.7}$ & 0.22$_{\pm0.06}$ & 105.31$_{\pm11.00}$ \\
\rowcolor[gray]{0.9} {\em Geomo} & \textbf{14.8$_{\pm1.1}$} & \textbf{39.0$_{\pm1.8}$} & \textbf{0.12$_{\pm0.10}$} & 218.74$_{\pm19.04}$ \\
\bottomrule
\end{tabular}
}
\end{table*}


Table~\ref{mae} reports energy-prediction MAE at three data scales (10k, 100k, and the full 460k setting) and on the in-distribution (ID) split as well as the three out-of-distribution splits, i.e., \(\mathrm{OOD}_{\mathrm{ads}}\), \(\mathrm{OOD}_{\mathrm{cat}}\), and \(\mathrm{OOD}_{\mathrm{both}}\). Across all three scales, \emph{Geomo} achieves the lowest MAE on each of these splits among all compared baselines, outperforming geometric models (CGCNN, DimeNet++, and EquiformerV2) as well as multimodal and text-only approaches (CatBERTa, GAP, and AdsMT). The consistent gains on \(\mathrm{OOD}_{\mathrm{both}}\) further demonstrate robust generalization to unseen adsorbate--catalyst combinations.
Complementary insights come from the coefficient of determination $R^2$ on the 460k setting, as shown in Figure \ref{R2460}(a) and Figure \ref{R2460}(b). {\em Geomo} achieves higher correlation on OOD$_{\text{ads}}$ and OOD$_{\text{both}}$, significantly higher than EquiformerV2 and AdsMT. This indicates that the learned motif-based representation captures the underlying physics of adsorption across diverse chemical spaces, rather than merely fitting the training distributions.

Figure~\ref{R2460}(c) compares inference speed and Top-1 hit rate, while Table II reports MAE. CatBERTa, CGCNN, and DimeNet++ exhibit fast inference but lower Top-1 hit rates; their MAE is roughly 0.2 eV higher than that of {\em Geomo} and EquiformerV2. Meanwhile, {\em Geomo} outperforms EquiformerV2 in both accuracy and efficiency: it delivers lower energy prediction error while being substantially faster than EquiformerV2. As a result, {\em Geomo} achieves a strong balance between predictive performance and computational cost.


\subsection{Motif-Based Fast Configuration Screening}

\textbf{Results and Analysis.}
We further perform adsorption configuration screening on the OC20Dense benchmark. To verify the generalization ability, all comparison models are directly evaluated on this dataset without task-specific retraining, enabling a rigorous assessment of screening performance under realistic practical scenarios. For each adsorbate-catalyst system, the model aims to select the structure corresponding to the global minimum of the predicted adsorption energy. 
We evaluate configuration selection performance using the following metrics:

\begin{enumerate}
    \item Top-$k$ Hit: Success rate that the true optimal configuration is among the model's top‑$k$ most stable predictions.

    



    \item Median Gap: the median of the energy differences $\Delta E$ across all systems, where each $\Delta E$ is the DFT energy penalty of the model-selected candidate relative to the true global minimum. It quantifies the typical selection error in a robust manner, being insensitive to outliers compared to the mean gap.

    \item Forward Samples/s: The inference throughput, i.e., the number of configurations processed per second, quantifying computational efficiency.
\end{enumerate}


As shown in Table~\ref{oc20dense_screening}, {\em Geomo} achieves the best Top-1 Hit, Top-3 Hit, and Median Gap, while maintaining competitive inference throughput. For the exact optimal configuration selection (Top-1 Hit), {\em Geomo} reaches 14.8\%, outperforming all baseline models by a substantial margin. The second-best method, EquiformerV2, attains 14.2\%, followed by Graphormer at 13.1\% and DimeNet++ at 12.8\%. A similar trend is observed for the Top-3 Hit metric, where {\em Geomo} scores 39.0\%, again surpassing the strongest baselines EquiformerV2 (36.6\%) and Graphormer (36.3\%). These results demonstrate that our method consistently identifies the DFT-optimal or near-optimal configurations more reliably than existing approaches.
Beyond hit rates, {\em Geomo} achieves a lower Median Gap of 0.12~eV, compared to the best baseline EquiformerV2 at 0.14~eV and Graphormer at 0.16~eV. This indicates that its chosen candidate is typically much closer in energy to the true optimum, with a median selection error of only 0.12~eV. This indicates that our method not only identifies the optimal configuration more frequently but also produces a more reliable prediction of the top candidates.

In terms of inference efficiency, lightweight models such as CatBERTa achieve the highest throughput (439.30 samples/s), while EquiformerV2 incurs a significantly higher computational cost (89.18 samples/s). Our {\em Geomo} operates at 218.74 samples/s, which is considerably faster than most geometric models and only moderately slower than the fastest text‑based baselines. This balance between high screening accuracy and competitive inference speed makes {\em Geomo} well suited for large‑scale, high‑throughput catalyst screening pipelines.

\begin{figure*}[htbp]
    \centering
    \vspace*{-0.3cm}
    \includegraphics[width=1.0\textwidth]{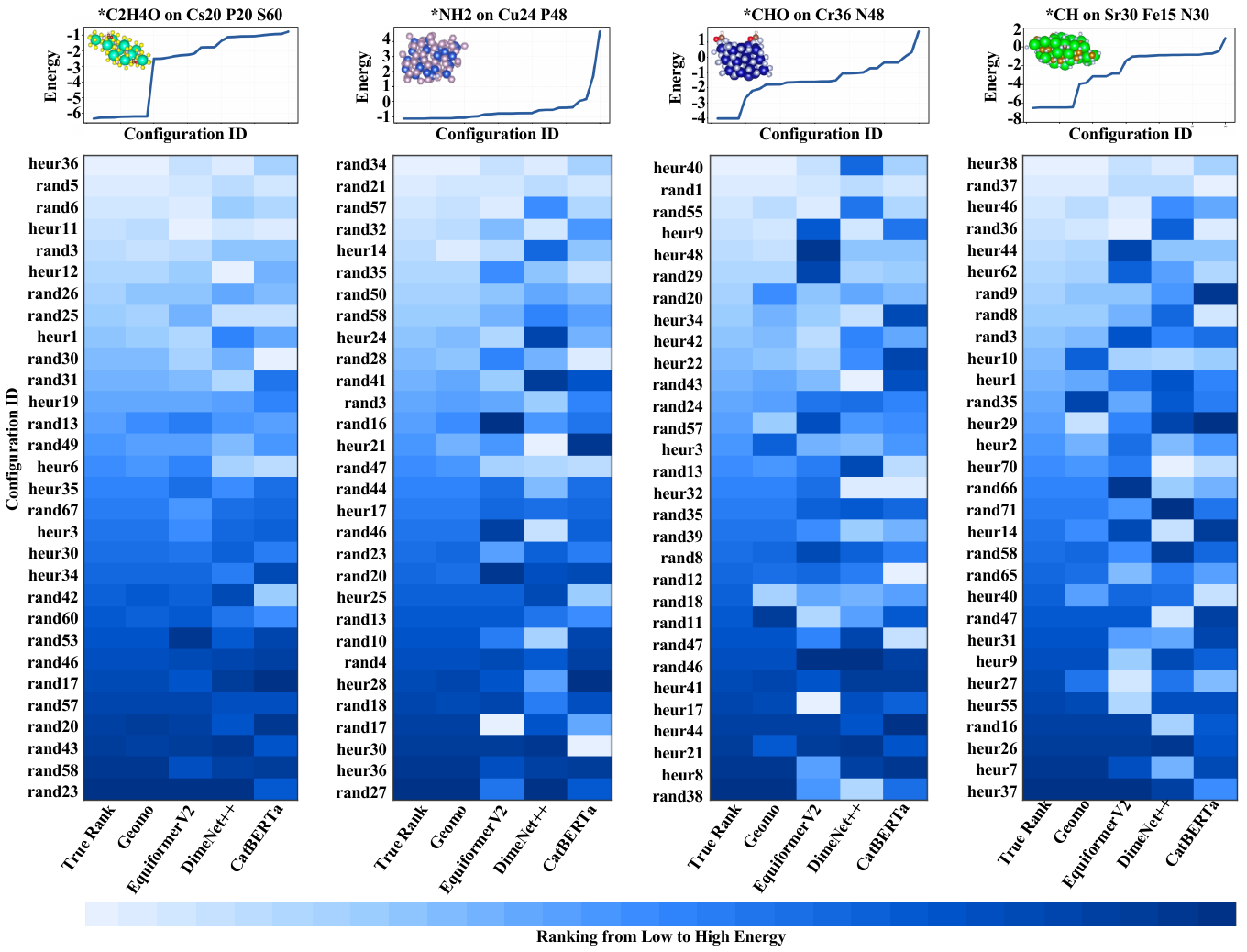}
    \vspace{-0.5cm}
    \caption{Energy distributions and ranking predictions for four representative adsorbate–catalyst systems (\ce{$\ast$C2H4O} on \ce{Cs20P20S60}, \ce{$\ast$NH2} on \ce{Cu24P48}, \ce{$\ast$CHO} on \ce{Cr36N48}, \ce{$\ast$CH} on \ce{Sr30Fe15N30}). Upper panel: four qualitatively different relaxed energy distributions. Lower panel: comparison of predicted DFT energy ranks by four methods.}
    \vspace{-0.3cm}
    \label{screeningmat}
\end{figure*}

\textbf{Ranking Visualization.}
To further validate the high-throughput performance, we screened adsorbate–catalyst systems with a larger number of configurations. Figure~\ref{screeningmat} presents four representative such systems: \ce{$\ast$C2H4O} on \ce{Cs20P20S60}, \ce{$\ast$NH2} on \ce{Cu24P48}, \ce{$\ast$CHO} on \ce{Cr36N48}, and \ce{$\ast$CH} on \ce{Sr30Fe15N30}. For each system, we include 30 distinct initial configurations that are generated via heuristic or random placement, and their relaxed energies are computed by DFT. The upper panel of the figure shows the relaxed energy distribution for each system, which exhibits four qualitatively different patterns: (a) a small population of low-energy configurations followed by a sharp transition to a large cluster of high-energy ones; (b) a large low-energy cluster gradually transitioning to a few high-energy outliers; (c) a very small number of low-energy configurations followed by a broad range of higher-energy structures; and (d) a continuous spectrum from low to high energies with comparable populations. These diverse energy landscapes provide a rigorous testbed for assessing model robustness across different distributional regimes.
The lower panel compares the true DFT energy ranks with the ranks predicted by four competing methods, including our \emph{Geomo}. 

Across all four systems and across both low- and high-energy regimes, {\em Geomo} consistently achieves rank predictions that align closely with the ground truth. In contrast, baseline methods often misorder configurations, especially near the low-energy tail, where accurate ranking is most critical for identifying promising candidates. These results demonstrate that \emph{Geomo} not only achieves high absolute energy accuracy but also preserves the relative ordering of configurations, which is essential for screening tasks where identifying the most stable structure is the primary objective.

\begin{figure*}[t!]
    \centering
    \vspace*{-0.3cm}
    \includegraphics[width=1.0\textwidth]{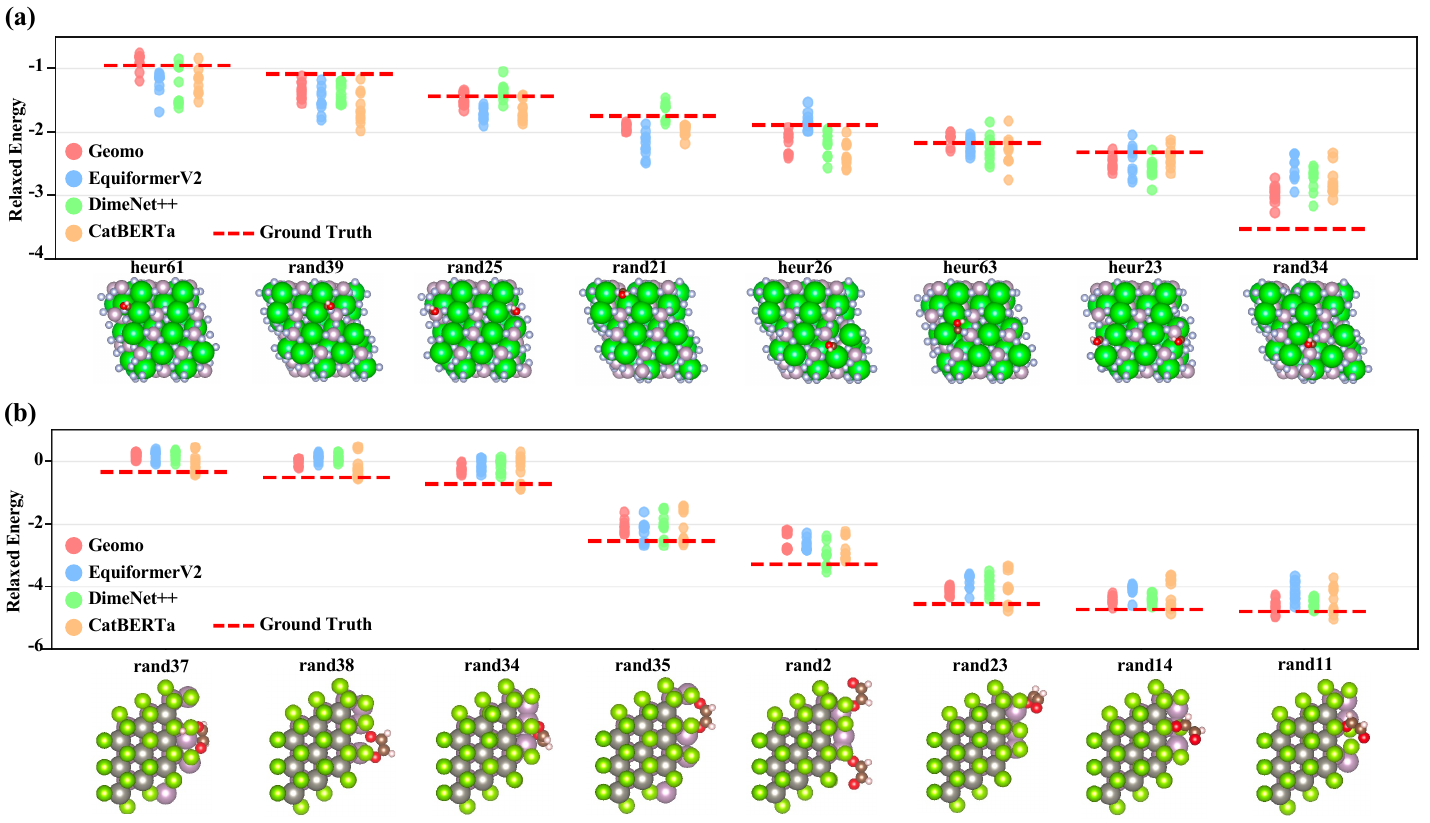}
    \vspace{-0.5cm}
    \caption{For each baseline model, we perform 10 independent predictions on two adsorbate–catalyst systems: \ce{$\ast$CHO} on \ce{Sr40Tc40} (a) and \ce{$\ast$C2H2O2} (b) on \ce{Mo18W9Se54}, and examine the distribution of the predicted adsorption energies.
    }
    \vspace{-0.3cm}
    \label{predictEnerCon}
\end{figure*}

\textbf{Screening Stability.}
Beyond qualitative ranking analysis, Figure~\ref{predictEnerCon} quantitatively evaluates the predicted energies. In this experiment, we train multiple models using different random seeds and apply them to the same set of configurations to assess screening stability. As shown in the figure, {\em Geomo} consistently predicts energies closer to the ground truth across all tested configurations. Moreover, the predictions from our method exhibit substantially lower variance across different training seeds compared to baseline methods, indicating superior stability and reduced sensitivity to initialization. These results demonstrate that {\em Geomo} not only achieves higher accuracy but also generalizes more reliably under realistic conditions where model training involves stochasticity.



\textbf{The Effect of Each Component.} In this part, we conduct ablation studies on the IS2RE task to quantify the individual contributions of \emph{Geomo}.
We first examine the adaptability of our framework by substituting the designed geometric encoder with other mainstream geometric encoders, including DimeNet++, EquiformerV2, Graphormer, and CGCNN, while keeping all other module configurations fixed. Figure~\ref{R2460}(c) presents the Top-1 hit rate and forward samples per second.
Among these geometric encoders, the proposed geometric encoder achieves the best trade-off between screening efficiency and Top-1 hit rate, justifying the rationality of using the designed geometric encoder.

We then conduct ablation experiments to validate the complementarity of the two modalities and the effectiveness of our proposed information bottleneck strategies. Two simplified variants are constructed for comparison: one purely relies on the designed geometric encoder to conduct energy regression without textual input (w/o Text), and the other removes geometric encoding branches and makes predictions merely based on chemical descriptions (w/o Geometry).
As shown in Figure~\ref{R2460}(d) and (e), the complete \emph{Geomo} framework achieves the minimal MAE, outperforming both single-modality variants. 
We then validate the effectiveness of the devised information bottleneck loss functions. As shown in Figure~\ref{R2460} (d) and (e), removing the information bottleneck for energy prediction (w/o YIB) or for text alignment (w/o MIB) does not significantly affect inference efficiency, but degrades the accuracy of relaxed energy prediction. This confirms the validity of our approach: extracting compact geometric motifs that align with textual features while retaining sufficient information for accurate energy prediction.


\subsection{Explainability Evaluation}

\textbf{Results and Analysis.} To assess whether our method identifies geometric motifs that align with chemists' understanding of adsorption behavior and to evaluate the faithfulness of its explanations, we construct a dedicated benchmark. Specifically, we define ground‑truth critical motifs using the well‑established locality principle of electronic structure: adsorption energy is dominated by the local chemical environment, and interactions beyond a radius of 4–6~\text{\AA} contribute negligibly to the total energy~\cite{norskov2009towards}. Following standard computational catalysis protocols~\cite{ock2023catalyst,tran2020methods}, we use the \texttt{Pymatgen} package~\cite{ong2013python} to partition the atomic environment into two physically distinct coordination shells (primary and secondary interaction regions, denoted $\mathcal{G}_{pri}$ and $\mathcal{G}_{sec}$) based solely on the distance $d_{ij}$ from each adsorbate atom $j$ to surface atoms. This benchmark is therefore independent of our model and reflects objective physical knowledge.

\begin{figure*}[t!]
    \centering
    \vspace*{-0.3cm}
    \includegraphics[width=1.0\textwidth]{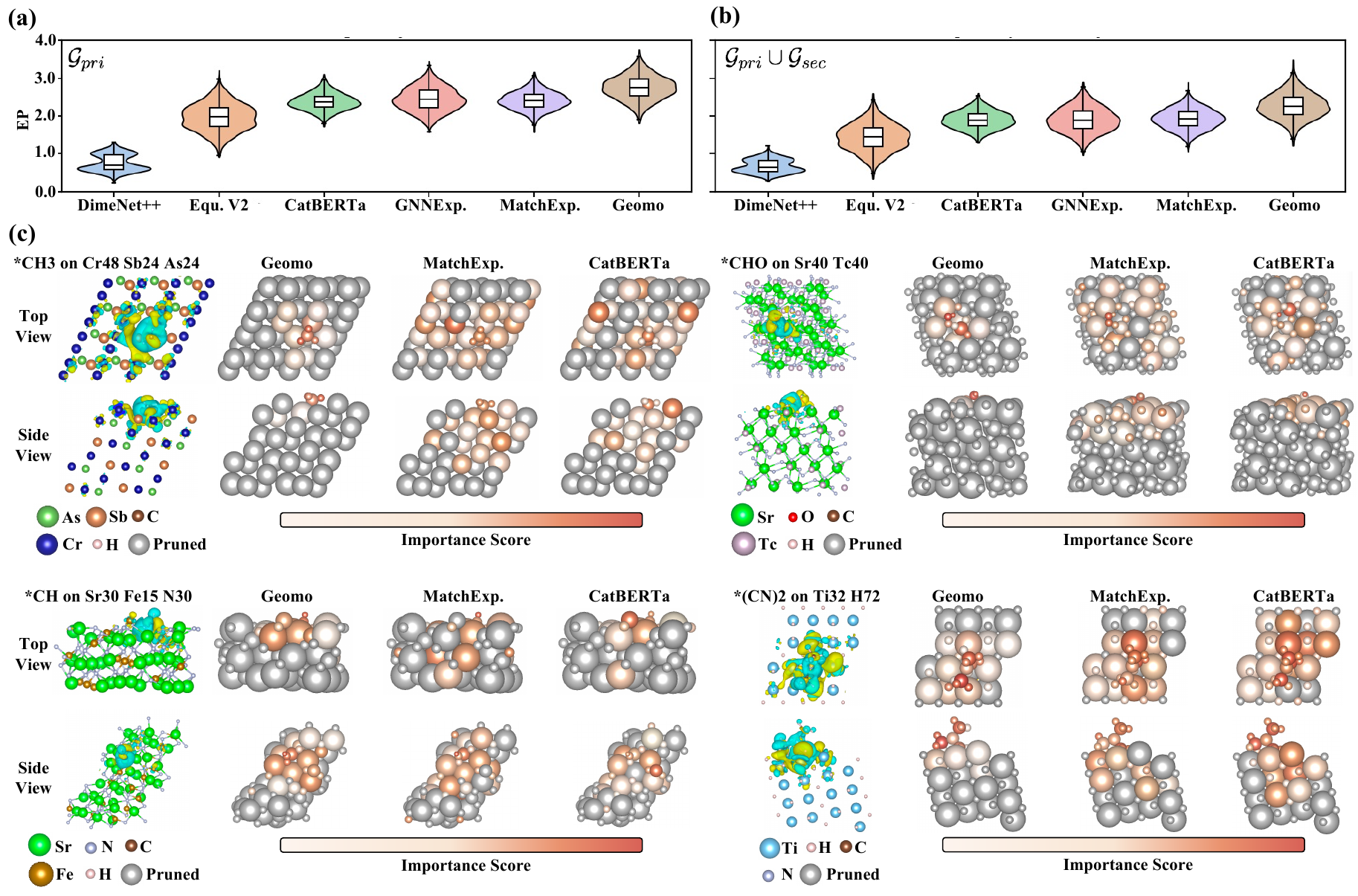}
    \caption{(a) EP distributions across ${\mathcal{G}}_{pri}$. (b) EP distributions across ${\mathcal{G}}_{pri} \cup {\mathcal{G}}_{sec}$. (c) Visualization of importance scores. The color gradient reflects model-derived importance from white to red, while gray color denotes unimportant atoms inferred by the models.
    }
    \label{expVis}
\end{figure*}

\begin{enumerate}
    \item Primary Interaction Region ($\mathcal{G}_{pri}$): This represents the first coordination shell where strong chemisorption (direct bonding) occurs. It is curated using a dynamic bonding threshold based on atomic sizes:
    $\mathcal{G}_{pri} = \{ \text{atom } i \mid \exists j \in \text{adsorbate}, d_{ij} \le (r_i + r_j) \times 1.2 \}$, where $r_i$ and $r_j$ are the covalent radii of the surface and adsorbate atoms, respectively.
    \item Secondary Interaction Region ($\mathcal{G}_{sec}$): This captures the second coordination shell, accounting for mid-range electronic perturbations (such as surface relaxation or d-band shifts) within the effective 4$\sim$6 \text{\AA} radius: $\mathcal{G}_{sec} = \{ \text{atom } i \mid \text{atom } i \notin \mathcal{G}_{pri} \text{ and } \exists k \in \mathcal{G}_{pri}, d_{ik} \le R_{cut} \}$,    where $R_{cut}$ is defined to encompass secondary neighbors contributing to the adsorption stability.
\end{enumerate}

Based on the annotated local structures, we introduce Explanation Plausibility (EP)~\cite{wang2026explainablemolecularpropertyprediction} to quantify how closely the model's explanations align with the ground‑truth geometric motifs. Additionally, we adopt Precision, Recall, and F1-score (P/R/F1) to measure the matching quality of the identified key motifs against the annotated primary and secondary interaction regions.

\begin{enumerate}
    \item EP: Given the importance scores $v$ over $N$ atoms, we apply a scale-invariant rank-squared Transformation $\hat{v}_i = [Rank(v_i) / (N-1)]^2$, where $Rank(\cdot)$ denotes the ascending rank. Let $\mathcal{T} \in \{\mathcal{G}_{pri}, \mathcal{G}_{pri} \cup \mathcal{G}_{sec}\}$ represent the benchmark set of atom indices and $\mathbb{I}(\cdot)$ be the indicator function. The mean transformed scores for the benchmark and background regions are denoted by $\bar{v}_{true} = {\textstyle \frac{\sum_{j=1}^N \hat{v}_j \cdot \mathbb{I}(j \in \mathcal{T})}{\sum_{j=1}^N \mathbb{I}(j \in \mathcal{T})}}$ and $\bar{v}_{false} = {\textstyle \frac{\sum_{j=1}^N \hat{v}_j \cdot \mathbb{I}(j \notin \mathcal{T})}{\sum_{j=1}^N \mathbb{I}(j \notin \mathcal{T})}}$, respectively. The $EP$ is defined as: $EP = \frac{\bar{v}_{true} - \bar{v}_{false}}{\bar{v}_{false}}$. The higher the $EP$ value, the more accurately the derived explanation is aligned with the annotations.
    
 \item P/R/F1: Given the benchmark set $\mathcal{T}$ and the key atom set $\mathcal{G}_{\text{local}}$, where $\mathcal{G}_{\text{local}}$ is either identified directly by \emph{Geomo} or constructed for baseline methods by selecting the top-$k$ atoms according to their importance scores ($k$ is the number of selected atoms matches that of \emph{Geomo}), we evaluate the identification quality as a set-matching problem. The P and R are denoted by $P = {\textstyle \frac{|\mathcal{G}_{m} \cap \mathcal{T}|}{|\mathcal{G}_{m}|}}$ and $R = {\textstyle \frac{|\mathcal{G}_{m} \cap \mathcal{T}|}{|\mathcal{T}|}}$, while the F1 is defined as their harmonic mean $F1 = 2 \cdot P \cdot R / (P + R)$. 
    \end{enumerate}

\begin{table*}[t]
\centering
\caption{Precision (P), Recall (R), and F1-score (F1) of different models trained on the 460k dataset for critical local region identification across ${\mathcal{G}}_{pri}$ and ${\mathcal{G}}_{pri} \cup {\mathcal{G}}_{sec}$. The highest result is \textbf{bold}.}
\label{expeva}
\setlength{\tabcolsep}{10pt}
\resizebox{\textwidth}{!}{
\small
\begin{tabular}{l ccc ccc}
\toprule
\multirow{4}{*}{Method} & \multicolumn{3}{c}{ $\mathcal{G}_{pri}$} & \multicolumn{3}{c}{$\mathcal{G}_{pri}  \cup \mathcal{G}_{sec}$} \\
\cmidrule(lr){2-4} \cmidrule(lr){5-7}
& P $\uparrow$ & R $\uparrow$ & F1 $\uparrow$ & P $\uparrow$ & R $\uparrow$ & F1 $\uparrow$ \\
\midrule
CGCNN (GradCAM)~\cite{selvaraju2017grad}        & 0.183$_{\pm0.002}$ & 0.350$_{\pm0.003}$ & 0.240$_{\pm0.002}$ & 0.461$_{\pm0.021}$ & 0.317$_{\pm0.035}$ & 0.376$_{\pm0.026}$  \\
DimeNet++ (GradCAM)~\cite{selvaraju2017grad}    & 0.278$_{\pm0.004}$ & 0.455$_{\pm0.004}$ & 0.345$_{\pm0.004}$ & 0.376$_{\pm0.002}$ & 0.271$_{\pm0.003}$ & 0.315$_{\pm0.002}$ \\
EquiformerV2 (GradCAM)~\cite{selvaraju2017grad} & 0.569$_{\pm0.004}$ & 0.874$_{\pm0.002}$ & 0.689$_{\pm0.003}$ & 0.770$_{\pm0.003}$ & 0.525$_{\pm0.003}$ & 0.625$_{\pm0.003}$ \\
CatBERTa (GradCAM)~\cite{selvaraju2017grad} & 0.506$_{\pm0.005}$ & 0.855$_{\pm0.003}$ & 0.630$_{\pm0.004}$ & 0.739$_{\pm0.003}$ & 0.510$_{\pm0.004}$ & 0.612$_{\pm0.003}$ \\
GNNExplainer~\cite{ying2019gnnexplainer}     & 0.574$_{\pm0.004}$ & 0.920$_{\pm0.003}$ & 0.707$_{\pm0.003}$ & 0.822$_{\pm0.002}$ & 0.579$_{\pm0.002}$ & 0.680$_{\pm0.002}$ \\
MatchExplainer~\cite{pmlrv202wu23jExplaining}     & 0.580$_{\pm0.003}$ & 0.911$_{\pm0.005}$ & 0.709$_{\pm0.004}$ & 0.834$_{\pm0.005}$ & 0.585$_{\pm0.002}$ & 0.688$_{\pm0.003}$ \\
\rowcolor[gray]{0.9} {\em Geomo} & \textbf{0.606$_{\pm0.004}$} & \textbf{0.984$_{\pm0.005}$} & \textbf{0.750$_{\pm0.004}$} & \textbf{0.899$_{\pm0.003}$} & \textbf{0.685$_{\pm0.002}$} & \textbf{0.778$_{\pm0.003}$} \\
\bottomrule
\end{tabular}
}
\end{table*}

Table~\ref{expeva} reports precision (P), recall (R), and F1-score for models trained on the full 460k dataset, using GradCAM-based attribution or dedicated explainers. 
Our method, {\em Geomo}, consistently achieves the highest scores across both region sets. Notably, on $\mathcal{G}_{pri}$, Geomo attains P=0.606, R=0.984, and F1=0.750, outperforming CatBERTa (P=0.506, R=0.855, F1=0.630). On the broader region set $\mathcal{G}_{pri}\cup\mathcal{G}_{sec}$, Geomo achieves P=0.899, R=0.685, and F1=0.778, again outperforming CatBERTa (P=0.739, R=0.510, F1=0.612).
This indicates that {\em Geomo} not only recalls most relevant atoms but also maintains tighter focus on truly critical sites, avoiding the scattered false positives common in gradient-based methods (e.g., CGCNN and DimeNet++).

Figure~\ref{expVis}(a) and Figure~\ref{expVis}(b) further quantify the EP values. A higher EP signifies that the model sharply concentrates its attention on the primary coordination environment rather than peripheral atoms. {\em Geomo} achieves the highest median EP of 2.41 on $\mathcal{G}_{pri}$, substantially outperforming GNN-based baselines and exceeding CatBERTa. When expanding to $\mathcal{G}_{pri} \cup \mathcal{G}_{sec}$, the EP remains superior, confirming that our framework systematically amplifies physically meaningful regions while suppressing chemically irrelevant ones.

\textbf{Explanation Visualization.}
To validate the accuracy and physical interpretability of our proposed machine learning method, we further investigated the consistency between the key structural motifs identified by our proposed model and the differential charge density.
The differential charge density of adsorbate-catalyst configurations reveals the electronic interaction mechanism between adsorbates and catalyst surfaces during the adsorption process, and we use the Quantum ESPRESSO package to obtain the electron cloud distribution regions.

Figure~\ref{expVis}(c) presents a visualized comparison of differential charge density distributions and the key structural motifs screened by different algorithms for various adsorbate–catalyst configurations. The results demonstrate that the proposed {\em Geomo} assigns high importance scores to atoms located exactly within the interfacial electron accumulation and depletion regions, including the immediate bonding environment and the secondary coordination sphere that dominate interfacial electronic interactions. By comparison, baseline methods frequently deviate from the valid charge density regions, either overemphasizing atomic sites far from the adsorption interface or failing to capture the critical surface atoms that dominate electronic redistribution. 

This distinct advantage indicates that our model can precisely localize the physically active regions consistent with differential charge density characteristics rather than relying on purely data-driven fitting. Overall, {\em Geomo} effectively identifies physically credible coordination motifs that closely match the electronic interaction regions revealed by charge density analysis, effectively bridging the gap between pure data-driven prediction and physical catalytic mechanisms, and further verifying the accuracy and practicality of the interpretable machine learning framework for catalytic adsorption analysis.

\textbf{Quantitative Correlation with Differential Charge Density.} To quantitatively assess whether the atoms identified as important by our model are physically grounded in electronic structure, we systematically compare the model-assigned importance scores with differential charge density. The overall experimental procedure is threefold: (i) each grid point from the first-principles differential charge density map is assigned to its nearest atom via Voronoi tessellation, such that each atom is associated with a corresponding set of grid points; (ii) for each atom, we extract two complementary descriptors from its assigned grid points to characterize the extent of electron cloud reorganization around that atom, which serves as a physics-based proxy for atomic importance; (iii) we compute the Spearman rank correlation between the model-predicted importance rankings and the rankings derived from each of the two charge-density descriptors. This analysis reveals whether the structural features relied upon by our model are consistent with the actual electronic-structure mechanisms underlying adsorption.

The differential charge density is computed using the Quantum ESPRESSO package, yielding a three-dimensional grid of $80 \times 96 \times 280 = 2{,}150{,}400$ grid points, each associated with a spatial coordinate $(x_g, y_g, z_g)$ and a differential charge density value $\Delta \rho_g$. To map this continuous grid information onto discrete atoms, we employ Voronoi tessellation: each grid point is assigned to the nearest atom based on Euclidean distance, partitioning the entire grid into $N$ disjoint regions, each corresponding to a specific atom $i$. We collected 100 different configurations of differential charge density data\footnote{The data are publicly available at https://github.com/fenghao2002/electron-density-difference-data-112-systems}.

\begin{figure*}[t!]
    \centering
    \vspace*{-0.3cm}
    \includegraphics[width=1.0\textwidth]{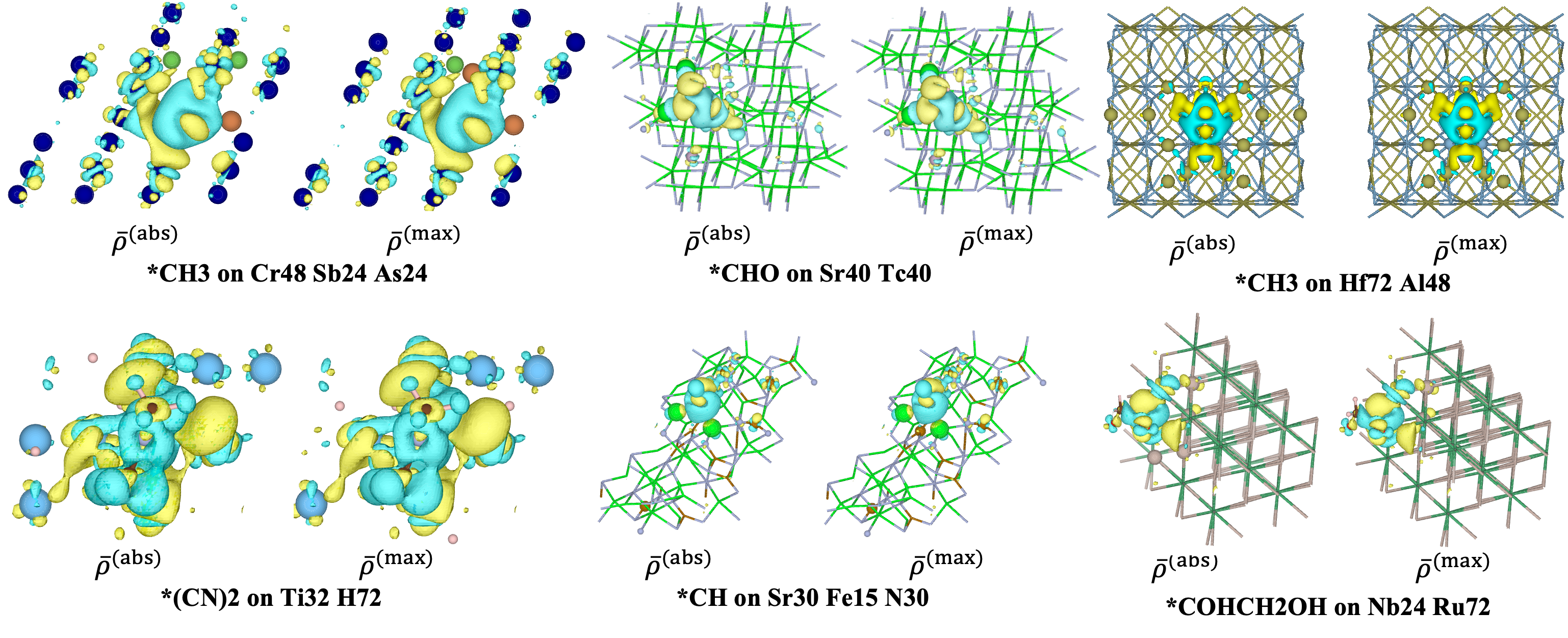}
    \vspace{-0.5cm}
\caption{The importance of atoms is quantitatively evaluated using the proposed integrated absolute charge density $\rho^{\text{(abs)}}$ and local peak charge density $\rho^{\text{(max)}}$ metrics. For $\ast$CH3 on Cr48--Sb24--As24 and $\ast$(CN)2 on Ti32--H72, the top-30 atoms with the highest $\rho^{\text{(abs)}}$ and $\rho^{\text{(max)}}$ values are displayed; for $\ast$CHO on Sr40--Tc40 and $\ast$CH on Sr30--Fe15--N30, the top-15 atoms with the highest $\rho^{\text{(abs)}}$ and $\rho^{\text{(max)}}$ values are shown; for $\ast$CH3 on Hf72--Al48 and $\ast$COHCH2OH on Nb24--Ru72, the top-20 and top-11 atoms values are shown, respectively. The important atoms identified by both metrics exhibit strong consistency with the differential charge density distributions.}
    \vspace{-0.3cm}
    \label{chargedensitydistributions}
\end{figure*}

\begin{table*}[htbp]
\centering
\caption{Spearman rank correlation coefficients between the true and method-identified integrated absolute charge density $\rho^{\text{(abs)}}$ and local peak charge density $\rho^{\text{(max)}}$}
\label{tab:model_comparison}
\resizebox{\textwidth}{!}{
\begin{tabular}{@{} l *{7}{c} @{}} 
\toprule
\textbf{} & 
\makecell{{CGCNN} \\(GradCAM)} & 
\makecell{{DimeNet++} \\(GradCAM)} & 
\makecell{{EquiformerV2} \\(GradCAM)} & 
\makecell{{CatBERTa} \\(GradCAM)} & 
{GNNExplainer} & 
{MatchExplainer} & 
{\em Geomo} \\
\midrule
$\rho^{\text{(abs)}}$ & 0.5881  & 0.6212 & 0.7990 & 0.8066   & 0.8041 & 0.8218 & \textbf{0.8737} \\
$\rho^{\text{(max)}}$ & 0.5766 & 0.6219 & 0.8038 & 0.7913 &0.8030 & 0.8255 & \textbf{0.8675} \\
\bottomrule
\end{tabular}
}
\end{table*}

We define two complementary atomic-level descriptors from the differential charge density. The first is the integrated absolute charge density:
\begin{equation}
\bar{\rho}_i^{\text{(abs)}} = \sum_{g \in \text{Voronoi}(i)} |\Delta \rho_g|,
\end{equation}
which sums the absolute values of all positive and negative differential charge densities within the Voronoi region of atom $i$. This descriptor quantifies the total amount of electron cloud reorganization around each atom induced by adsorption, where positive and negative values correspond to charge accumulation and depletion, respectively—both of which are treated as contributing signals.

The second descriptor is the local peak charge density:
\begin{equation}
\bar{\rho}_i^{\text{(max)}} = \max_{g \in \text{Voronoi}(i)} |\Delta \rho_g|.
\end{equation}
In contrast to the integrated descriptor, which captures overall reorganization, this descriptor focuses on the strongest local charge density variation within each atom's Voronoi region. It is designed to capture sharp features arising from localized interactions such as directional bonding or orbital hybridization at the adsorption interface.

For each of the two descriptors, we rank all atoms in descending order of $\bar{\rho}_i$, yielding a charge-density-based atomic ranking. Separately, we rank all atoms in descending order of the model-assigned importance scores $s_i$. We then compute the Spearman rank correlation coefficient between the two full rankings across all $N$ atoms:
\begin{equation}
\rho^{\text{(abs)}} = \frac{ \sum_{i=1}^{N} (r_i - \bar{r})(q_i - \bar{q}) }{ \sqrt{ \sum_{i=1}^{N} (r_i - \bar{r})^2 \sum_{i=1}^{N} (q_i - \bar{q})^2 } },
\end{equation}
where $r_i$ and $q_i$ are the ranks of atom $i$ according to the charge-density descriptor and the model, respectively, and $\bar{r}$, $\bar{q}$ are their mean values. The same calculation is applied using the second descriptor $\bar{\rho}_i^{\text{(max)}}$ to obtain $\rho^{\text{(max)}}$. 
A high correlation indicates that the atoms deemed important by our model coincide with those exhibiting the strongest differential charge density features, thereby providing direct electronic-structure evidence that the identified motifs are physically meaningful rather than merely statistically correlated.

We validate that the proposed integrated absolute charge density metric $\rho^{\text{(abs)}}$ and local peak charge density metric $\rho^{\text{(max)}}$ can reflect the differential electron cloud density and quantitatively assess atomic importance. Specifically, we conduct visualizations, as shown in Fig.~\ref{chargedensitydistributions}, for $\ast$CH$_3$ on Cr$_{48}$Sb$_{24}$As$_{24}$ and $\ast$(CN)$_2$ on Ti$_{32}$H$_{72}$, the top-30 atoms with the highest $\rho^{\text{(abs)}}$ and $\rho^{\text{(max)}}$ are displayed, while for $\ast$CHO on Sr$_{40}$Tc$_{40}$ and $\ast$CH on Sr$_{30}$Fe$_{15}$N$_{30}$, the top-15 most atoms are visualized. For $\ast$CH3 on Hf72--Al48 and $\ast$COHCH2OH on Nb24--Ru72, the top-20 and top-11 atoms values are shown, respectively. The results indicate that the atoms identified by both metrics are in good agreement with the differential electron cloud density. The atomic importance scores obtained from these two metrics can be used as the ground truth for representing the intrinsic significance of the atoms.

Based on the Spearman rank correlation coefficients reported in Table~\ref{tab:model_comparison}, we systematically compare the atomic importance rankings identified by various graph neural network architectures and explanation methods against two physics-based benchmarks derived from first-principles differential charge density, namely $\rho^{\text{(abs)}}$ and $\rho^{\text{(max)}}$. Overall, all methods yield positive correlation coefficients, indicating a consistent alignment between the atoms attended by the models and the extent of electron cloud reorganization revealed by density functional theory calculations.

Among different model architectures, the GradCAM-based interpretations exhibit clear performance stratification. CGCNN and DimeNet++ achieve correlation coefficients of approximately 0.59 and 0.62, respectively, which are relatively modest compared to more advanced architectures. In contrast, EquiformerV2 and CatBERTa attain substantially higher coefficients around 0.80, marking an improvement over CGCNN and DimeNet++. 
For dedicated explainability methods, GNNExplainer and MatchExplainer yield correlation coefficients ranging from 0.80 to 0.82, which are comparable to those of EquiformerV2 and CatBERTa. Our proposed \emph{Geomo} framework achieves the highest correlation among all compared methods, with coefficients of 0.8737 and 0.8675 on the $\rho^{\text{(abs)}}$ and $\rho^{\text{(max)}}$ benchmarks, respectively. This outperforms the next best method, MatchExplainer by approximately 0.05, suggesting that \emph{Geomo} exhibits the strongest agreement with the electronic-structure-based atomic importance rankings.

Furthermore, when comparing the two benchmarks, the correlation coefficients for each method differ only marginally between $\rho^{\text{(abs)}}$ and $\rho^{\text{(max)}}$. This consistency indicates that the important atoms identified by the models are simultaneously reflected in both the total integrated charge reorganization and the local peak variations.
In summary, the results in Table~\ref{tab:model_comparison} provide a quantitative, electronic-structure-grounded comparison of atomic importance identification across baseline models and explanation techniques, offering empirical evidence for assessing the physical plausibility of model interpretations in heterogeneous catalysis.

\section{Conclusion}

We presented \emph{Geomo}, a multimodal information-bottleneck framework for learning physically meaningful geometric motifs in molecule–surface systems. Unlike black-box surrogates that merely fit target values, \emph{Geomo} formulates motif discovery as an information-theoretic problem, aligning 3D geometry with textual expert knowledge and adsorption energy to extract compact motifs that are both predictive and physically meaningful. These motifs are not post-hoc explanations; they are embedded as structured physical priors in the model's reasoning pathway, jointly improving accuracy and transparency.
On the representative task of catalytic configuration energy prediction, \emph{Geomo} achieves competitive performance in energy prediction across diverse out-of-distribution splits and attains 14.8\% Top-1 and 39.0\% Top-3 hit rates in high-throughput screening. The configurations it selects incur a median energy penalty of only 0.12 eV above the DFT global minimum. Moreover, the discovered motifs quantitatively correlate with differential charge density maps, confirming that the model recovers physically meaningful structure--property relationships.
These results suggest a route from black-box surrogate modeling to interpretable, physics-aware multimodal representation learning for scientific discovery.

\bibliographystyle{IEEEtran}
\bibliography{references}

\end{document}